\documentclass[a4paper,11pt]{article}
\pdfoutput=1
\usepackage{jcappub} 
\usepackage{lineno}

\usepackage[utf8]{inputenc}
\usepackage{enumitem}
\usepackage{graphicx}
\usepackage{hyperref}
\usepackage{mathtools}
\usepackage{comment}
\usepackage{amsmath}
\usepackage{amssymb}
\usepackage{lipsum} 
\usepackage{float}
\usepackage{cleveref}
\usepackage[caption=false]{subfig}
\usepackage{placeins}
\usepackage{xcolor}
\usepackage{float}

\arxivnumber{2512.18935} 
\title{Weak Lensing Mass Calibration of the ACT DR5 Galaxy Clusters with the DES Year 3 Weak Lensing Data}
\author[a,1]{T. Shin,\note{Corresponding author.}}
\author[b]{E. J. Baxter,}  
\author[c,d]{E. Lee,} 
\author[d]{N. Battaglia,}
\author[e]{A. Alarcon,}
\author[f]{A. Amon,} 
\author[g]{M. Becker,}
\author[c]{G. Bernstein,} 
\author[h]{J. R. Bond,}
\author[a]{A. Campos,}
\author[i,j]{C. Chang,} 
\author[k]{R. Chen,}
\author[l]{A. Choi,}
\author[m]{J. DeRose,}
\author[i,j,n]{S. Dodelson,}
\author[o]{C. Doux,} 
\author[p,f]{J. Dunkley,}
\author[q]{J. Elvin-Poole,}
\author[r]{J. H. Esteves,}
\author[s]{S. Everett,}
\author[t]{A. Fert\'e,}
\author[j]{M. Gatti,}
\author[u]{S. Grandis,}
\author[v]{D. Gruen,}
\author[w]{I. Harrison,}
\author[x]{J. C. Hill,}
\author[y,z]{M. Hilton,}
\author[c]{M. Jarvis,}
\author[aa]{N. MacCrann,}
\author[f,k,t]{J. McCullough,}
\author[z,ab]{K. Moodley,}
\author[e]{T. Mroczkowski,}
\author[f]{J. Myles,}
\author[ac]{A. Navarro Alsina,}
\author[ad]{A. Nicola,}
\author[ae]{L. Page,}
\author[af]{S. Pandey,}
\author[ag]{J. Prat,}
\author[ah]{M. Raveri,}
\author[ai,k]{B. Ried Guachalla,}
\author[ad]{R. P. Rollins,}
\author[aj,ak]{C. Sanchez,}
\author[j]{L. F. Secco,}
\author[m]{E. Sheldon,}
\author[al]{C. Sif\'on,}
\author[am]{M. Troxel,}
\author[an]{I. Tutusaus,}
\author[ao]{A. von der Linden,}
\author[ap]{E. Wollack,}
\author[am]{B. Yin,}
\author[aq,ar]{M. Aguena,}
\author[n]{S. S. Allam,}
\author[as]{O. Alves,}
\author[at]{F. Andrade-Oliveira,}
\author[au]{D. Bacon,}
\author[v]{S. Bocquet,}
\author[av]{D. Brooks,}
\author[aw]{R. Camilleri,}
\author[ax,ar,ay]{A. Carnero Rosell,}
\author[ak]{J. Carretero,}
\author[az,e]{F. J. Castander,}
\author[ba,aq,bb]{M. Costanzi,}
\author[ar]{L. da Costa,}
\author[bc]{M. E. da Silva Pereira,}
\author[aw]{T. Davis,}
\author[bd]{J. De Vicente,}
\author[be]{S. Desai,}
\author[n]{B. Flaugher,}
\author[i,n,j]{J. Frieman,}
\author[bf]{J. Garcia-Bellido,}
\author[n]{G. Gutierrez,}
\author[aw]{S. Hinton,}
\author[bg]{D. L. Hollowood,}
\author[as]{D. Huterer,}
\author[bh]{D. James,}
\author[bi]{S. Lee,}
\author[bj]{J. Marshall,}
\author[o]{J. Mena-Fernández,}
\author[bk,bl]{F. Menanteau,}
\author[bm,ak]{R. Miquel,}
\author[v]{J. Mohr,}
\author[bn,bo]{J. Muir,}
\author[bp,bq]{R. Ogando,}
\author[k,t]{A. Plazas Malagón,}
\author[bd,br]{A. Porredon,}
\author[bs]{K. Romer,}
\author[bd]{E. Sanchez,}
\author[bd,at]{D. Sanchez Cid,}
\author[bd]{I. Sevilla,}
\author[bt]{M. Smith,}
\author[at]{M. Soares-Santos,}
\author[bu]{E. Suchyta,}
\author[bk]{M. Swanson,}
\author[i]{C. To,}
\author[bv,bw]{N. Weaverdyck,}
\author[bx,v]{J. Weller,}
\affiliation[a]{Department of Physics and Astronomy, Carnegie Mellon University,\\ 5000 Forbes Ave, Pittsburgh, PA 15213, USA}
\affiliation[b]{Institute for Astronomy, University of Hawaii,\\ 640 North A‘ohōkū Place, Hilo, HI 96720, USA}
\affiliation[c]{Department of Physics \& Astronomy, University of Pennsylvania,\\ 209 S 33rd St., Philadelphia, PA 19104, USA}
\affiliation[d]{Department of Astronomy, Cornell University,\\ Ithaca, NY 14853, USA}
\affiliation[e]{Institute of Space Sciences (CSIC-ICE),\\ Carrer de Can Magrans, s/n, 08193 Cerdanyola del Vall\`es, Barcelona, Spain}
\affiliation[f]{Department of Astrophysical Sciences, Princeton University,\\ Peyton Hall, Princeton, NJ 08544, USA}
\affiliation[g]{Argonne National Laboratory,\\ 9700 South Cass Avenue, Lemont, IL 60439, USA}
\affiliation[h]{Canadian Institute for Theoretical Astrophysics,\\ 60 St. George Street, 
University of Toronto, Toronto, ON, M5S 3H8, Canada}
\affiliation[i]{Department of Astronomy and Astrophysics, University of Chicago,\\ Chicago, IL 60637, USA}
\affiliation[j]{Kavli Institute for Cosmological Physics, University of Chicago,\\ Chicago, IL 60637, USA}
\affiliation[k]{Kavli Institute for Particle Astrophysics and Cosmology,\\ 382 Via Pueblo Mall Stanford, CA 94305-4060, USA}
\affiliation[l]{NASA Goddard Space Flight Center,\\ 8800 Greenbelt Rd, Greenbelt, MD 20771, USA}
\affiliation[m]{Brookhaven National Laboratory,\\ Bldg 510, Upton, NY 11973, USA}
\affiliation[n]{Fermi National Accelerator Laboratory,\\ P. O. Box 500, Batavia, IL 60510, USA}
\affiliation[o]{Universit\'e Grenoble Alpes, CNRS, LPSC-IN2P3,\\ 38000 Grenoble, France}
\affiliation[p]{Joseph Henry Laboratories of Physics, Jadwin Hall, Princeton University,\\ Princeton, NJ, USA 08544}
\affiliation[q]{Department of Physics and Astronomy, University of Waterloo,\\ 200 University Ave W, Waterloo, ON N2L 3G1, Canada}
\affiliation[r]{Laboratory for Particle Physics and Cosmology, Harvard University,\\ Cambridge, MA 02138, USA}
\affiliation[s]{California Institute of Technology,\\ 1200 East California Blvd, MC 249-17, Pasadena, CA 91125, USA}
\affiliation[t]{SLAC National Accelerator Laboratory,\\ Menlo Park, CA 94025, USA}
\affiliation[u]{Universit\"at Innsbruck, Institut f\"ur Astro- und Teilchenphysik,\\ Technikerstr. 25/8, 6020 Innsbruck, Austria}
\affiliation[v]{University Observatory, LMU Faculty of Physics,\\ Scheinerstr. 1, 81679 Munich, Germany}
\affiliation[w]{School of Physics and Astronomy, Cardiff University,\\ CF24 3AA, UK}
\affiliation[x]{Department of Physics, Columbia University,\\ New York, NY, USA 10027}
\affiliation[y]{Wits Centre for Astrophysics, School of Physics, University of the Witwatersrand,\\ Private Bag 3, 2050, Johannesburg, South Africa}
\affiliation[z]{Astrophysics Research Centre, University of KwaZulu-Natal, Westville Campus,\\ Durban 4041, South Africa}
\affiliation[aa]{Department of Applied Mathematics and Theoretical Physics, University of Cambridge,\\ Cambridge CB3 0WA, UK}
\affiliation[ab]{School of Mathematics, Statistics \& Computer Science, University of KwaZulu-Natal, Westville Campus,\\ Durban4041, South Africa}
\affiliation[ac]{Instituto de F\'isica Gleb Wataghin, Universidade Estadual de Campinas,\\ 13083-859, Campinas, SP, Brazil}
\affiliation[ad]{Jodrell Bank Centre for Astrophysics, Department of Physics and Astronomy, The University of Manchester,\\ Manchester M13 9PL, UK}
\affiliation[ae]{Department of Physics, Princeton University,\\ Princeton, NJ 08544, USA}
\affiliation[af]{Department of Physics and Astronomy, Johns Hopkins University,\\ Baltimore, MD 21218, USA}
\affiliation[ag]{Nordita, KTH Royal Institute of Technology and Stockholm University,\\ Hannes Alfv\'ens v\"ag 12, SE-10691 Stockholm, Sweden}
\affiliation[ah]{Department of Physics, University of Genova and INFN,\\ Via Dodecaneso 33, 16146, Genova, Italy}
\affiliation[ai]{Department of Physics, Stanford University, Stanford, CA 94305-4085, USA}
\affiliation[aj]{Departament de F\'{\i}sica, Universitat Aut\`{o}noma de Barcelona (UAB),\\ 08193 Bellaterra, Barcelona, Spain}
\affiliation[ak]{Institut de F\'{\i}sica d'Altes Energies (IFAE), The Barcelona Institute of Science and Technology, Campus UAB,\\ 08193 Bellaterra (Barcelona) Spain}
\affiliation[al]{Instituto de F\'isica, Pontificia Universidad Cat\'olica de Valpara\'iso,\\ Casilla 4059, Valpara\'iso, Chile}
\affiliation[am]{Department of Physics, Duke University,\\ Durham, NC 27708, USA}
\affiliation[an]{Institut de Recherche en Astrophysique et Plan\'etologie (IRAP), Universit\'e de Toulouse, CNRS, UPS, CNES,\\ 14 Av. Edouard Belin, 31400 Toulouse, France}
\affiliation[ao]{Department of Physics and Astronomy, Stony Brook University,\\ Stony Brook, NY 11794, USA}
\affiliation[ap]{NASA Goddard Space Flight Center,\\ Greenbelt, MD 20771, USA}
\affiliation[aq]{INAF-Osservatorio Astronomico di Trieste,\\ via G. B. Tiepolo 11, I-34143 Trieste, Italy}
\affiliation[ar]{Laborat\'orio Interinstitucional de e-Astronomia - LIneA,\\ Av. Pastor Martin Luther King Jr, 126 Del Castilho, Nova Am\'erica Offices, Torre 3000/sala 817 CEP: 20765-000, Brazil}
\affiliation[as]{Department of Physics, University of Michigan,\\ Ann Arbor, MI 48109, USA}
\affiliation[at]{Physik-Institut, University of Zürich,\\ Winterthurerstrasse 190, CH-8057 Zürich, Switzerland}
\affiliation[au]{Institute of Cosmology and Gravitation, University of Portsmouth,\\ Portsmouth, PO1 3FX, UK}
\affiliation[av]{Department of Physics \& Astronomy, University College London,\\ Gower Street, London, WC1E 6BT, UK}
\affiliation[aw]{School of Mathematics and Physics, University of Queensland,\\  Brisbane, QLD 4072, Australia}
\affiliation[ax]{Instituto de Astrofisica de Canarias,\\ E-38205 La Laguna, Tenerife, Spain}
\affiliation[ay]{Universidad de La Laguna, Dpto. Astrofísica,\\ E-38206 La Laguna, Tenerife, Spain}
\affiliation[az]{Institut d'Estudis Espacials de Catalunya (IEEC),\\ 08034 Barcelona, Spain}
\affiliation[ba]{Astronomy Unit, Department of Physics, University of Trieste,\\ via Tiepolo 11, I-34131 Trieste, Italy}
\affiliation[bb]{Institute for Fundamental Physics of the Universe,\\ Via Beirut 2, 34014 Trieste, Italy}
\affiliation[bc]{Hamburger Sternwarte, Universit\"{a}t Hamburg,\\ Gojenbergsweg 112, 21029 Hamburg, Germany}
\affiliation[bd]{Centro de Investigaciones Energ\'eticas, Medioambientales y Tecnol\'ogicas (CIEMAT),\\ Madrid, Spain}
\affiliation[be]{Department of Physics, IIT Hyderabad,\\ Kandi, Telangana 502285, India}
\affiliation[bf]{Instituto de Fisica Teorica UAM/CSIC, Universidad Autonoma de Madrid,\\ 28049 Madrid, Spain}
\affiliation[bg]{Santa Cruz Institute for Particle Physics,\\ Santa Cruz, CA 95064, USA}
\affiliation[bh]{Center for Astrophysics $\vert$ Harvard \& Smithsonian,\\ 60 Garden Street, Cambridge, MA 02138, USA}
\affiliation[bi]{Jet Propulsion Laboratory, California Institute of Technology,\\ 4800 Oak Grove Dr., Pasadena, CA 91109, USA}
\affiliation[bj]{George P. and Cynthia Woods Mitchell Institute for Fundamental Physics and Astronomy, and Department of Physics and Astronomy, Texas A\&M University,\\ College Station, TX 77843,  USA}
\affiliation[bk]{Center for Astrophysical Surveys, National Center for Supercomputing Applications,\\ 1205 West Clark St., Urbana, IL 61801, USA}
\affiliation[bl]{Department of Astronomy, University of Illinois at Urbana-Champaign,\\ 1002 W. Green Street, Urbana, IL 61801, USA}
\affiliation[bm]{Instituci\'o Catalana de Recerca i Estudis Avan\c{c}ats,\\ E-08010 Barcelona, Spain}
\affiliation[bn]{Department of Physics, University of Cincinnati,\\ Cincinnati, Ohio 45221, USA}
\affiliation[bo]{Perimeter Institute for Theoretical Physics,\\ 31 Caroline St. North, Waterloo, ON N2L 2Y5, Canada}
\affiliation[bp]{Centro de Tecnologia da Informa\c{c}\~ao Renato Archer,\\ Campinas, SP, Brazil - 13069-901}
\affiliation[bq]{Observat\'orio Nacional,\\ Rio de Janeiro, RJ, Brazil - 20921-400}
\affiliation[br]{Ruhr University Bochum, Faculty of Physics and Astronomy, Astronomical Institute, German Centre for Cosmological Lensing,\\ 44780 Bochum, Germany}
\affiliation[bs]{Department of Physics and Astronomy, Pevensey Building, University of Sussex,\\ Brighton, BN1 9QH, UK}
\affiliation[bt]{Physics Department, Lancaster University,\\ Lancaster, LA1 4YB, UK}
\affiliation[bu]{Computer Science and Mathematics Division, Oak Ridge National Laboratory,\\ Oak Ridge, TN 37831}
\affiliation[bv]{Berkeley Center for Cosmological Physics, Department of Physics, University of California,\\ Berkeley, CA 94720, US}
\affiliation[bw]{Lawrence Berkeley National Laboratory,\\ 1 Cyclotron Road, Berkeley, CA 94720, USA}
\affiliation[bx]{Max Planck Institute for Extraterrestrial Physics,\\ Giessenbachstrasse, 85748 Garching, Germany}
\emailAdd{taehyeos@andrew.cmu.edu}

\abstract{We use weak gravitational lensing measurements from Year 3 Dark Energy Survey data to calibrate the masses of 443 galaxy clusters selected via the Sunyaev-Zel'dovich effect from Atacama Cosmology Telescope Data Release 5 maps of the cosmic microwave background. 
We incorporate redshift and SZ measurements for individual clusters into a hierarchical model for the stacked lensing signals and perform Bayesian analyses to constrain the hydrostatic mass bias of the clusters. 
Our treatment of systematic uncertainties includes a prescription for measuring and accounting for the weak lensing boost factor, consideration of a miscentering effect, as well as marginalization over uncertainties in the source galaxy photometric redshift distributions and shear calibration. 
The resultant constraints on the normalization of the mass-observable relation have a precision of approximately 7\%, with the mean WL halo mass of $M_{\rm 500c} = 5.4 \times 10^{14} M_{\odot}$. 
We measure the bias between the true cluster mass and the mass estimated from the SZ signal based on an X-ray--calibrated scaling relation assuming hydrostatic equilibrium, to be $1-b = 0.74^{+0.06}_{-0.05}$ over the full sample. 
When splitting the clusters into high ($z$=0.43-0.70) and low ($z$=0.15-0.43) redshift bins, we measure $1-b = 0.58^{+0.06}_{-0.05}$ and $0.81^{+0.08}_{-0.06}$, respectively. 
When introducing additional freedom in redshift and mass to the hydrostatic bias model, we find that $1-b$ decreases with redshift (with the power law of $-1.8^{+0.5}_{-0.6}$, 99.95\% confidence), consistent with findings from other recent studies, while we do not find any significant trend in mass. 
We also demonstrate that our result is robust against various systematics such as a scale cut, priors on baryonic and miscentering parameters, and degree of scatter in mass-observable relation. 
The weak-lensing mass calibration presented in this study will be a useful tool for using the ACT clusters as probes of astrophysics, and as a step towards using their abundance as a cosmological probe.}

\begin{document}

\begin{nolinenumbers}
\vspace*{-\headsep}\vspace*{\headheight}
\footnotesize \hfill FERMILAB-PUB-25-0859-V\\
\vspace*{-\headsep}\vspace*{\headheight}
\footnotesize \hfill DES-2025-0951
\end{nolinenumbers}

\maketitle
\flushbottom

\section{Introduction}

Galaxy clusters form at the highest density peaks of the early Universe, evolving through hierarchical mergers. 
Consequently, their mass and abundance could serve as a powerful cosmological probe, constraining the matter content ($\Omega_{\rm m}$) and the fluctuation of the matter field ($\sigma_8$) \cite{Haiman2001,Dodelson2016}. 
Therefore, the most important task for cluster cosmology is to accurately and precisely constrain the mass of the cluster sample of interest to retrieve an unbiased cosmology.
The mass of a galaxy cluster also controls other halo properties such as the mass and temperature of the intra-cluster medium (ICM), the number of galaxies hosted by the cluster halo (richness) and its size. 
These ``observable'' quantities, being tightly correlated to the mass of the cluster, provide proxies for the cluster mass which in general cannot be directly used to characterize clusters from observations.
Therefore, in order to perform an unbiased cosmological analysis using galaxy clusters, the mass-observable relation must be calibrated in advance or jointly. 

To date, the majority of galaxy clusters have been detected through the optical and near-infrared emission of the galaxies they host.
Several wide-field optical imaging surveys, including the Dark Energy Survey \cite[DES,][]{DES}, the Kilo-Degree Survey \cite[KiDS,][]{KiDS}, the Hyper Suprime-Cam Survey \cite[HSC,][]{HSC} have amassed samples of tens of thousands of galaxy clusters \cite{CAMIRA, McClintock:2019,DESY3Cluster,Maturi2025}.  
However, characterizing optical cluster selection and the relation between optical cluster observables (e.g., the cluster richness) and the underlying cluster mass has proven to be challenging, in that the scatter between richness and mass could be large, as well as dependent on a number of cluster and survey properties, which are not yet fully understood (e.g., projection effect and detection aperture effect; see, for example, refs. \cite{Zu2017,Busch2017,DESY1CLUSTER,Sunayama2020,Wu2022,Sunayama2023}).

Another efficient way to detect galaxy clusters is via the Sunyaev-Zel'dovich (SZ) effect \cite{SZ1972}, caused by the inverse Compton scattering of photons from the Cosmic Microwave Background (CMB) with the cluster ICM, which transports the energy of the CMB photons to higher frequencies therefore leaving characteristic signatures on the CMB map.  
The SZ signal is known to correlate more tightly with the cluster mass, and to be less susceptible to selection biases than optical selection \cite[e.g.,][]{daSilva2004,Nagai2006,Battaglia2012,Krause2012,Kay2012,Hoekstra2012,Sifon2013}.   
Moreover, unlike the optical selection, SZ cluster detection can be extended to the redshift of formation in principl, since the SZ signal is largely independent of redshift.  
Until recently, the cluster cosmology with SZ-selected clusters had been limited by the small size of cluster samples \cite[e.g.,][]{Reichardt2013,Hasselfield2013}.
However, high sensitivity and wide-area CMB surveys from the Atacama Cosmology Telescope \cite[ACT,][]{ACT1,ACT2}, the South Pole Telescope \cite[SPT,][]{SPT} and the {\it Planck} mission \cite{Planck} have expanded these samples to $\mathcal{O}(10^4)$ clusters \cite{Bleem2015,PlanckClusters,Bleem2020,Hilton2021,Klein2023,Bleem2024,ACTDR6clusters}, enabling robust cosmological analyses using SZ-selected clusters \cite{Bocquet2019,Bocquet2023,Bocquet2024,Lee2024,Aymerich2025}.
In the future, CMB surveys such as Simons Observatory \cite[SO,][]{SO} will further expand the number of SZ-selected clusters \cite{Abitbol2025}.
In order to fully harness the potential of these large cluster samples, a precise and unbiased prescription of the cluster mass calibration must come first.

Under the assumption of hydrostatic equilibrium (HSE) of the ICM within clusters, the ICM-related cluster observables such as the SZ signal ($Y_{\rm SZ}$ which is proportional to integrated pressure) and X-ray photon count could be related theoretically to the cluster mass. 
Specifically, for the galaxy clusters from tSZ observations, the hydrostatic mass of the clusters could be derived using the X-ray observations through the amplitude of the tSZ signal anchored on the corresponding X-ray profiles. 
Nevertheless, existence of significant non-thermal pressure support within the ICM (e.g., turbulent motion; refs. \cite{Shi2016,Shi2018,Green2020}) breaks the HSE assumption, causing biases in the cluster mass estimates.
This discrepancy between the hydrostatic mass estimate and true cluster mass is referred to as the hydrostatic mass bias ($b$), and is parameterized by $M_{\rm  HSE} = (1-b) M_{\rm true}$.  
Consequently, the uncertainty on this hydrostatic mass bias, or equivalently, the normalization of the mass-observable relations, represents a dominant source of systematic uncertainty for cluster cosmology with SZ and X-ray selected clusters (e.g., refs. \cite{Bocquet2024,Ghirardini2024} for recent results).

Weak gravitational lensing (WL), through its ability to directly probe the total projected matter field, provides a powerful way to unbiasedly measure cluster masses and thus to calculate the hydrostatic mass bias, or equivalently calibrate the mass-observable relation, of a given set of observed clusters (e.g., see ref. \cite{Bartelmann:2010} for a fuller review on the gravitational lensing). It is also free from the assumption on the connection between dark and visible matter. The mass calibration of galaxy clusters using gravitational lensing may be performed in a stacked method \cite[e.g.,][]{Simet2017,Leauthaud2017,Melchior2017,McClintock:2019,Robertson2024,Shirasaki2024,KiDS_clusters}, where the signals from individual clusters are combined into one set of data vectors, or in a hierarchical method, where the signal from each cluster is used individually in a Bayesian hierarchical manner\cite[e.g.,][]{Applegate2014,Bocquet2024,Grandis2024lensing,Chiu2025}. On the other hand, WL measurements are susceptible to several important sources of systematic uncertainty. For example, using background galaxies to measure the lensing signal requires constraints on the galaxy redshift distribution, and biases and uncertainties in inference of these distributions could easily propagate into the mass calibration \cite[e.g.,][]{McClintock:2019, Grandis2021, Myles2021}. A related issue is the so-called `boost factor', the contamination from the cluster member galaxies that do not carry any lensing signal leaking into the background source galaxy sample due to the photometric redshift uncertainties \cite[e.g.,][]{Sheldon2004,Applegate2014,Gruen2014,Hoekstra2015,Simet2017,Melchior2017,Leauthaud2017,Medezinski2018source,Varga2019,Grandis2024lensing}. Another challenge for cluster mass calibration is miscentering: offsets in the assumed cluster centers with respect to the true cluster centers could bias the resultant mass calibration \cite[e.g.,][]{Zhang2019miscentering,Kelly2023,Sommer2024,Ding2025}. In addition, biases in the galaxy shape measurement algorithm propagate into the lensing measurements, typically inducing a few percent shift \cite[e.g.,][]{MacCrann2022}. 

In this work, we present a weak lensing mass calibration analysis of SZ-selected clusters from the 5th data release of ACT (ACT DR5). Our analysis relies on WL measurements with the DES year-3 (DESY3) galaxy catalogs. In effect, we measure the stacked reduced shear profiles around the ACT DR5 clusters using the DESY3 source galaxy shapes to constrain the cluster masses, therefore the hydrostatic mass bias. This measurement is enabled by the $\sim$ 4600 square degrees of sky overlap between the DES and ACT surveys.

We use the DES weak lensing measurements to constrain the absolute mass scale of the ACT clusters and to constrain the relationship between the SZ observable and the cluster mass. Several aspects of our analysis are notable: (1) we use the entire tomographic source redshift bins defined by the DESY3 3$\times$2-point cosmology analysis \cite{DES3x2} instead of a custom selection of source galaxies to recycle the already calibrated photo-z and shear biases; note that our framework of using the entire tomographic source redshift bins was also proposed and applied by refs. \cite{Grandis2024lensing,Bocquet2023, Bocquet2024}, (2) we account for known important systematics in a fully Bayesian manner (i.e. parameterizing and marginalizing over them), (3) we incorporate the redshifts and SZ signals for individual clusters when building our model for the stacked lensing signal, (4) we incorporate a prescription for baryonic effects on the cluster mass distribution based on Cromer et al. \cite{Cromer2022}, which avoids calculating the WL mass bias from realistic cosmological simulations (see, e.g., ref. \cite{Grandis2024lensing} for the calibration of the WL mass bias using simulations), and (5) we include the evolution of the hydrostatic mass bias in redshift and its dependence on cluster mass.

Our analysis takes a fully forward model for the tangential shear signal around the clusters as a function of the cluster-centric radial angle: we model the shear profiles of individual clusters that are then stacked to obtain the final stacked tangential shear model, while all the systematic biases such as the multiplicative shear bias, the bias in the redshift distribution and the boost factor correction are applied at the modeling stage. This approach assures that our measured data vectors are independent of our model parameters.

The paper is organized as follows. In section~\ref{sec:data} we describe the data sets that we use in this work: galaxy clusters from ACT, and galaxy shapes and redshifts measured from DES imaging.  In section~\ref{sec:measurement} we describe our primary measurements, the stacked tangential shear profile around the ACT clusters, and the estimation of boost factors, for a set of cluster bins.  
In section~\ref{sec:model} we describe our models for these two measurements, including a detailed treatment of several potential sources of systematic error.  
We describe the calculation of the likelihood for our measurements given our model in section~\ref{sec:likelihood}. Our main results are presented in section~\ref{sec:result} and we conclude in section~\ref{sec:discussion}. Throughout the paper, we assume a flat $\Lambda$CDM cosmology with $\Omega_{\rm m}=0.3075$, $\Omega_{\rm b}=0.0486$ and $H_0=67.74 {\rm km}s^{-1}{\rm Mpc}^{-1}$. Throughout the paper, we will use $M$ to refer to $M_{500c}$, the mass within the radius that encloses the mean density of $500$ times the critical density ($R_{\rm 500c}$).

\section{Data}
\label{sec:data}

\begin{figure}
    \centering
    \includegraphics[width=0.99\columnwidth]{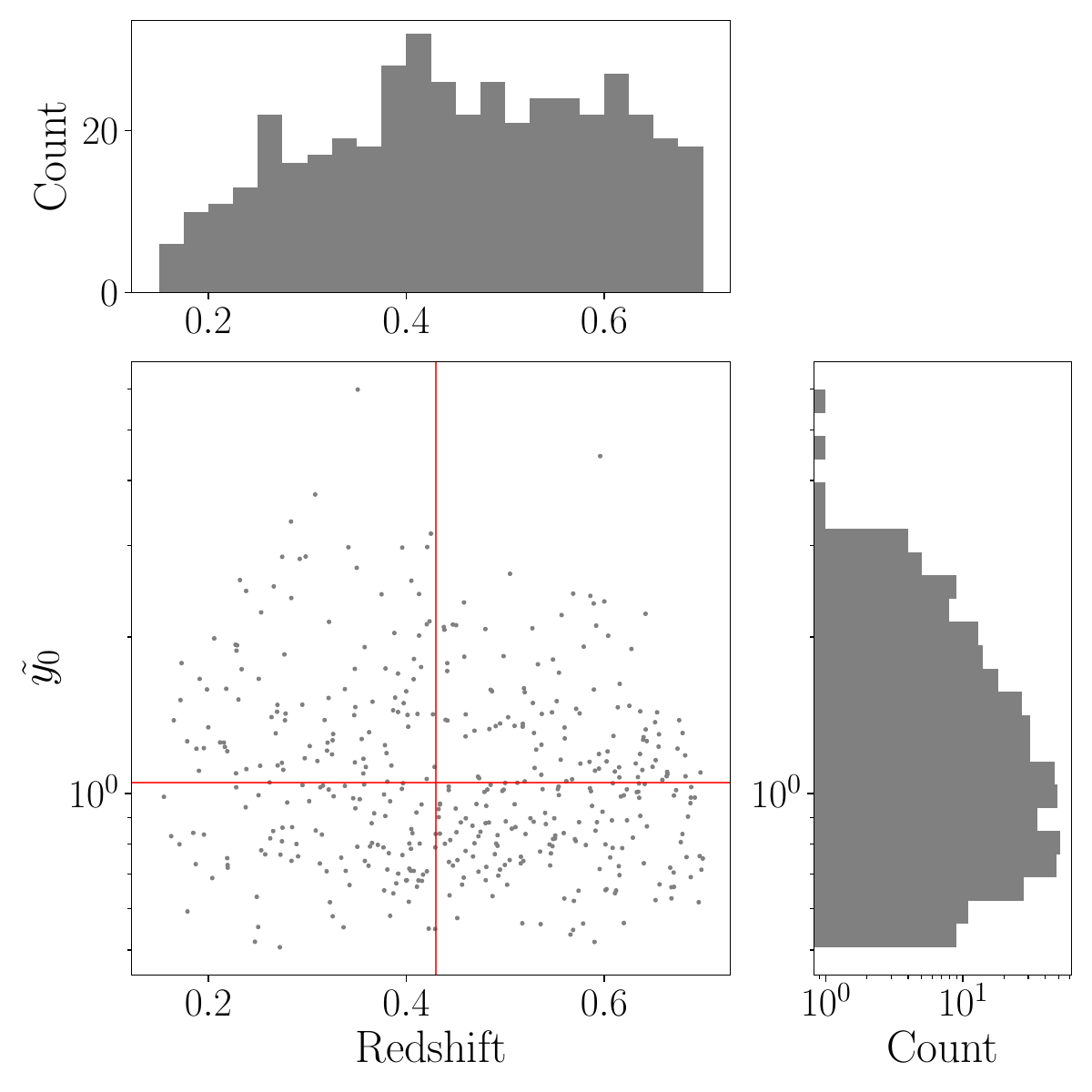}
    \caption{The distribution of the 443 Compton-$y$ parameters ($y_{2.4}$) and the redshifts of the cluster sample used in this study. The red lines demarcate our binning of the cluster sample.
    }
    \label{fig:dist_yz}
\end{figure}

\subsection{Galaxy Clusters from the ACT DR5}
\label{sec:data_act}

We use a sample of SZ-selected clusters from the ACT Data Release 5 \cite[ACT DR5]{Hilton2021}, which comprises 4195 galaxy clusters above the signal-to-noise ratio (SNR) of 4, with 13,221 square degrees of survey area. The clusters are detected with the matched-filter method applied to the 98 and 150 GHz maps constructed from the observations during 2008 to 2018 \cite{Naess2020}. We refer the reader to Hilton et al. \cite{Hilton2021} for details on this catalog, including optical confirmation and estimations of photometric (and/or spectroscopic) redshifts for the clusters.

We select the clusters that lie in the sky region covered by the Dark Energy Survey Year-3 (DESY3) shape catalog\footnote{Hereafter, by "DESY3" data, we mean the data collected from the first 3 years of DES observations.} which spans approximately 4,100 square degrees out of $\sim$5,000 square degrees of the total survey footprint (see more details below in section~\ref{sec:data_shear}) among which approximately 4,600 square degrees overlap with the ACT DR5 region (e.g., see Figure 2 of \cite{Mallaby-Kay2023} for the visualization of the overlapping footprint).
We further select the SZ clusters by following the recommendation of Hilton et al. \cite{Hilton2021} to apply a cut of ${\rm SNR}_{2.4}$\footnote{The SNR of the Compton-$y$ signal measured with a filter with the size of 2.4' ($y_{2.4})$, where the Compton-$y$ signal is proportional to the integrated electron pressure along the line-of-sight.}$>5.5$ cut, which minimizes the false detection rate and is appropriate for cosmological analyses (see their section 3.3 and figure 15).
We also remove those for which the centers of the clusters are not covered by the shape catalog, which is crucial for the WL mass calibration. In addition, we restrict our cluster selection to the redshift range of $z \in [0.15,0.70]$ to minimize the overlap between the lens and the source redshift distribution (see figure~\ref{fig:dist_source}) while preserving as much lensing power as possible. Our final sample thereby comprises $443$ galaxy clusters.  

Since per-cluster SNRs of the lensing measurement are low and it is difficult to model cluster-by-cluster profile variations such as triaxiality, concentration and large-scale projection \cite[e.g.,][]{Mandelbaum2008,Becker2011,Herbonnet2019,Herbonnet2022,Zhang2023}, we stack the lensing measurements over several clusters of comparable Compton-$y$ signal and redshift. 
To this end, we divide the cluster sample into two redshift bins ($z \in$ [0.15, 0.43] and $z \in$ [0.43, 0.7]) and two Compton-$y$ bins ($\tilde{y}_0 \in$ [0.0, 1.05] and $\tilde{y}_0 \in$ [1.05, 6.0]).\footnote{$\tilde{y}_0 = y_{2.4}/10^{-4}$} Our bin choice ensures that the SNRs of the weak lensing measurements for each bin are similar. We show the redshift and $\tilde{y}_0$ distribution of our cluster sample in figure~\ref{fig:dist_yz}. The red lines in the figure indicate the edges of our bins.

The catalog provides the mass ($M_{\rm 500c}$) of each cluster derived from the SZ signal with the hydrostatic equilibrium assumption, which we call $M_{\rm SZ}$. This mass is calculated from the SZ scaling relation from ref. \cite{Arnaud2010} which is calibrated using X-ray observations, as described in refs. \cite{Hasselfield2013,Hilton2021}. We build our hydrostatic mass bias model upon this $M_{\rm SZ}$, rather than assuming a new scaling relation based on an SZ observable such as $\tilde{y}_0$ or SNR$_{2.4}$(see section~\ref{sec:hs_bias} and \ref{sec:likelihood} for details).

\subsection{DESY3 Galaxy shape catalog}
\label{sec:data_shear}

DES is a multi-band photometric survey covering $\sim$5,000 square degrees of the South Galactic Cap. With the 570-megapixel Dark Energy Camera \cite{DECam} mounted on the Cerro Tololo Inter-American Observatory 4m Blanco telescope in Chile, it primarily produced images in \textit{grizY} filters. For our weak-lensing analysis, we make use of the galaxy shape catalog made from the first three years of the data \cite[DESY3,][]{Gatti2021} with the \textsc{Metacalibration} algorithm \cite{Huff2017,Sheldon2017}. \textsc{Metacalibration} measures the galaxy shapes, $\boldsymbol{\rm e}$, in the \textit{riz} bands, which is related to the shear through the response matrix $\mathcal{R}$\footnote{The response of the galaxy shape to the applied shear, both of which are spin-2 field, resulting in a 2$\times$2 matrix.}:
\begin{equation}
    \langle \boldsymbol{\gamma} \rangle  = \langle \mathcal{R} \rangle^{-1} \langle \boldsymbol{\rm e} \rangle.
\end{equation}
In addition to the shear response, the average galaxy shapes also depend on the specific selection one makes on the galaxy sample. We call that the selection response and denote it by $\mathcal{R}_{\rm s}$. 
The total response for a galaxy $i$ then becomes $\mathcal{R} = \mathcal{R}^i + \mathcal{R}_{\rm s}$. Note that the selection response is defined for the whole galaxy sample, and not for each galaxy. For a more detailed description of the method, we refer the readers to the above studies.

The calibration of the measured shear is performed by running a set of image simulations, as described in ref. \cite{MacCrann2022}. Throughout our analysis we use the same source galaxy selections as the DESY3 cosmology analyses \cite[e.g.,][]{DES3x2}.  We therefore adopt the values and uncertainties on the multiplicative shear bias parameters, $m_\gamma$, provided in that work (see further discussion in section~\ref{sec:sys_uncertainty}).  

\subsection{Photometric redshifts of source galaxies}
\label{sec:data_photoz}

\begin{figure}
    \centering
    \includegraphics[width=0.99\columnwidth]{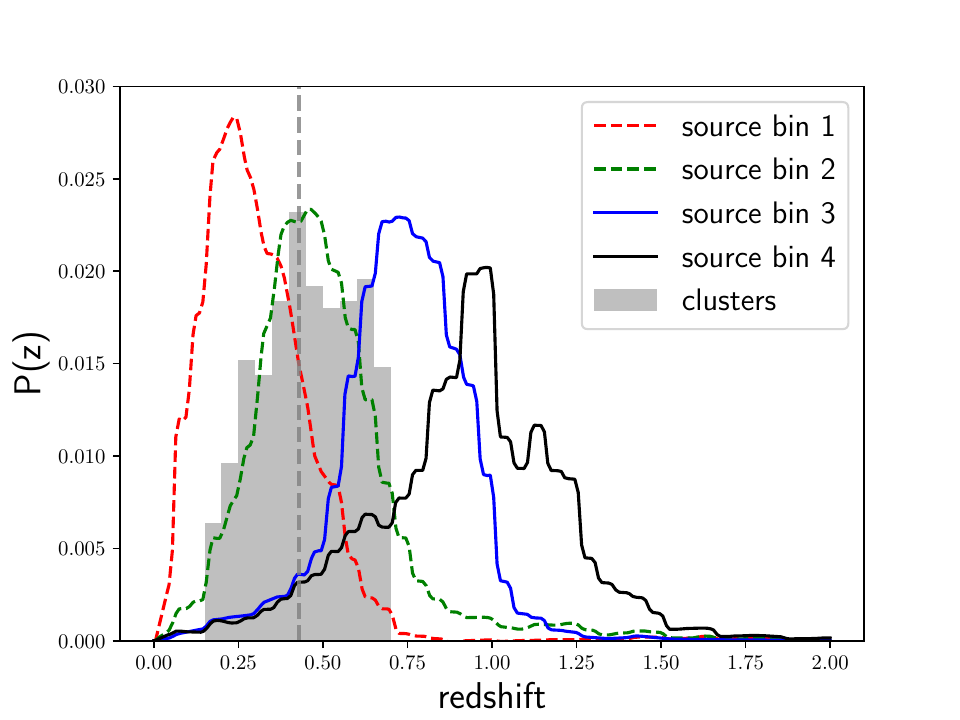}
    \caption{The redshift distributions of the galaxy clusters (gray) and source galaxies (other curves) considered in this work. The vertical gray line at $z=0.43$ shows the location where we split the clusters into two redshift bins. We exclude the first two source redshift bins (dashed) as majority of the galaxies within those bins are at or in front of our cluster sample, having no lensing signal.}
    \label{fig:dist_source}
\end{figure}

Our weak-lensing measurements use the same source galaxy redshift choice as the DES 3x2pt cosmology analysis \cite{DES3x2}. The redshift distributions, $n(z_{\rm s})$, of these galaxies were estimated using the Self-Organizing Map $P(z)$ (SOMPZ) algorithm described in ref. \cite{Myles2021}. The source galaxies were divided into 4 redshift bins (see figure~\ref{fig:dist_source}), for which we adopt the same priors for the biases in $n(z_{\rm s})$ thereof \cite{MacCrann2022}.
We refer readers to the references above for details of binning and calibration of the source galaxy redshift distributions. The details of the redshift bias model are discussed below in section~\ref{sec:sys_uncertainty}.

For the calculation of the boost factor (the contamination from the cluster member galaxies into our weak-lensing signal, see section~\ref{sec:measurement_boost} and \ref{sec:model_boost}), we use the photometric redshift estimates of the individual source galaxies obtained by the DNF algorithm \cite{SevillaNoarbe2021,DNF}.
It is because the SOMPZ method is optimized for the estimation of $n(z_{\rm s})$ (the population distribution) making the redshift estimation of individual galaxies highly uncertain, therefore not optimal for the boost factor calculation. 

\section{Measurements}
\label{sec:measurement}

\subsection{Tangential shear profiles}
\label{sec:measurement_gammat}

\begin{figure}
    \centering
    \includegraphics[width=0.99\columnwidth]{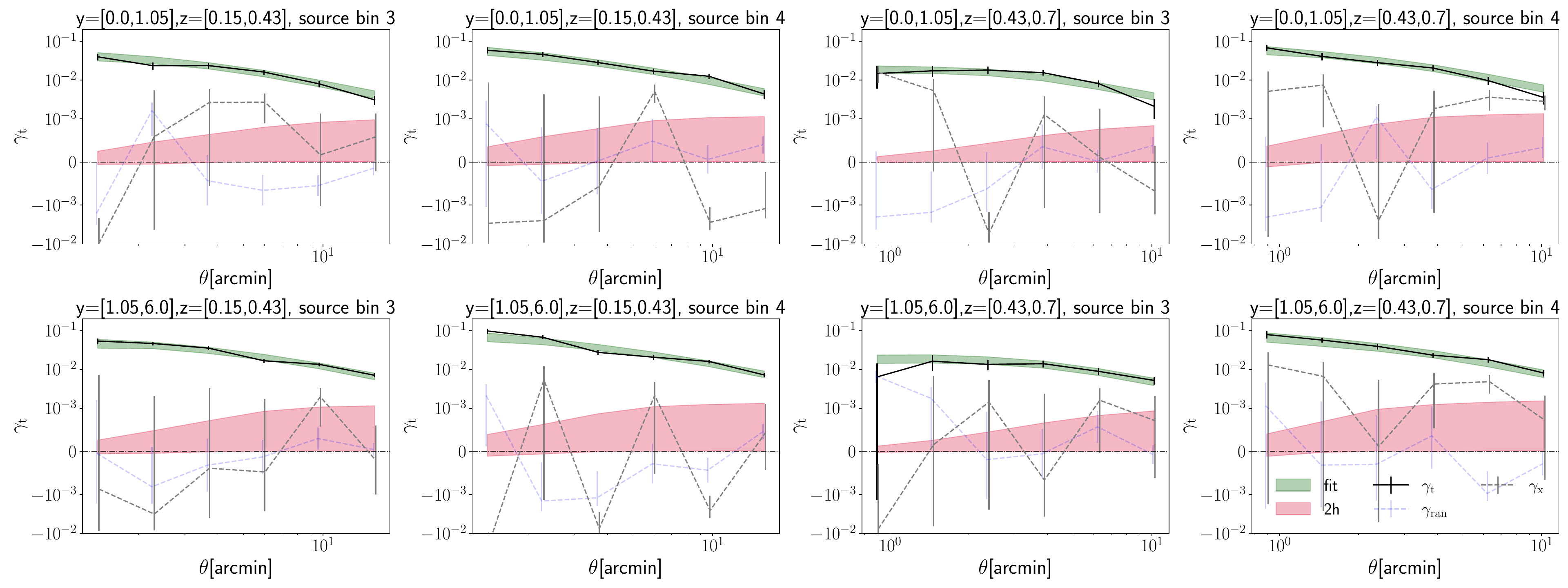}
    \includegraphics[width=0.99\columnwidth]{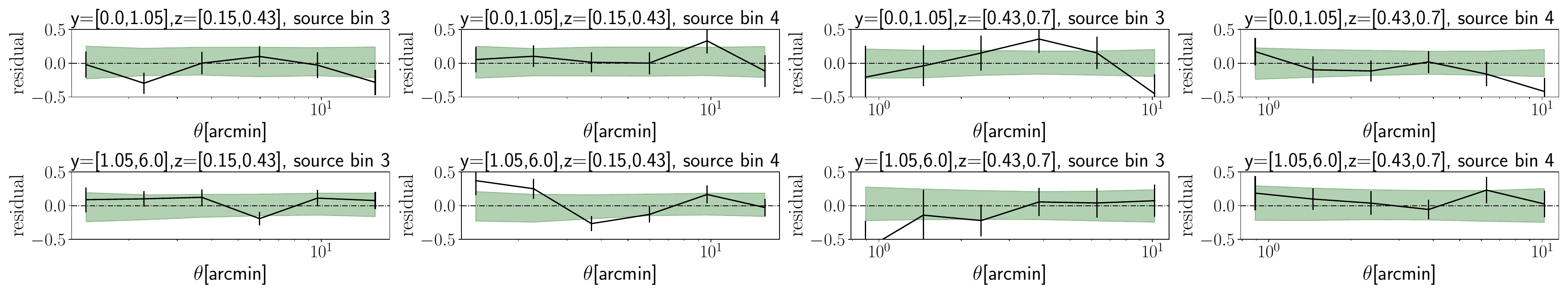}
    \caption{\textit{Top}: The tangential shear measurements around our cluster sample are shown with the black points with errorbars. The correlation of the same cluster sample with the cross-component of the shear is shown with the grey points with errorbars; the cross-shear component is expected to be consistent with zero in the absence of systematics. Also, the tangential shear measurements around random points are shown in light blue. The green bands represent the 68\% confidence ranges on the model fit to the data and the red bands those for the 2-halo contribution (we do not plot 1-halo components since they are indistinguishable from the total fit). The left (right) two columns correspond to the low (high) redshift cluster sample, while the top (bottom) panels correspond to the low (high) Compton-$y$ cluster sample.\\
    \textit{Bottom}: The corresponding fractional residuals with respect to the 68\% confidence range on the model fit, with the same plotting scheme as the top panel.}
    \label{fig:gammat_fit}
\end{figure}

The impact of gravitational lensing by the mass overdensity around galaxy clusters is to induce a tangential shear on the images of background galaxies. While individual clusters may not be exactly spherically symmetric, a spherically symmetric model is a good approximation to the average mass profile in each cluster bin, since the clusters are oriented randomly on the sky.  We therefore treat the stacked tangential shear profiles of the ACT galaxy clusters in bins of redshift and $\tilde{y}_0$ as our main lensing observable (see section~\ref{sec:data_act} for the binning scheme).  We describe below how we measure the weak lensing profiles for each of these bins.

The tangential component of the shear field ($\gamma_{\rm t}$) is given by 
\begin{equation}
    \gamma_t = -\gamma_1 \cos 2\phi - \gamma_2 \sin 2\phi, 
\end{equation}
where $\gamma_1$ and $\gamma_2$ are shear components in a Cartesian coordinate system, and $\phi$ is the position angle of the source galaxy relative to the \textit{x}-axis directing towards the east. 

In the presence of a cluster, the tangential shear with respect to the cluster center is then related to the surface mass density ($\Sigma$) of the cluster through the critical surface density ($\Sigma_{\rm crit}$):
\begin{equation}
     \bar{\gamma}_t(\theta) = \frac{\Delta \Sigma (\theta)}{\Sigma_{\rm crit}(z_l,z_s)},
\end{equation}
where $z_l$ ($z_s$) is the redshift to the lens (source) galaxy and $\theta$ the separation angle between the halo center and the source position.
Here, 
\begin{equation}
    \Sigma_{\rm crit}(z_l,z_s) = \frac{c^2}{4\pi G} \frac{D_{{\rm A},s}}{D_{{\rm A},l} D_{{\rm A},ls}}
\end{equation}
and
\begin{equation}
\label{eq:dsigma}
    \Delta \Sigma (\theta) = \langle \Sigma (<\theta) \rangle -\Sigma(\theta),
\end{equation}
where $D_{{\rm A},s}$ is the angular diameter distance to the source galaxy, $D_{{\rm A},l}$ that to the lens, $D_{{\rm A},ls}$ that between the source galaxy and the lens, and $\langle \Sigma (<\theta) \rangle$ is the mean surface density within the halo-centric angular separation $\theta$. 

In practice, the lensing effect also includes isotropic magnification of galaxy images (due to convergence, $\kappa$). Therefore, we observe the reduced shear, $\boldsymbol{g}_t$, rather than the true shear, $\gamma_t$.  These are related by
\begin{equation}
    \boldsymbol{g}_t = \frac{\gamma_t}{1 - \kappa},
\end{equation}
where $\kappa = \Sigma/\Sigma_{\rm crit}$.
Taking into account this $1-\kappa$ factor in the tangential shear model is especially important in the central regions of clusters where the surface density is high. 

We measure the (reduced) tangential shear around the cluster sample averaged within the redshift and $\tilde{y}_0$ bins (section~\ref{sec:data_act}), treating each source galaxy redshift bin (section~\ref{sec:data_photoz}) separately. For some source galaxy redshift bins, the majority of source galaxies are at lower redshifts than most of the clusters.  
For these bin combinations, the lensing signal is expected to be small, and we therefore remove them from our analysis.  As a result, we exclude the first two source redshift bins (see figure~\ref{fig:dist_source}).

With the measured galaxy shapes and the responses (section~\ref{sec:data_shear}), our estimator for the tangential shear given a cluster bin $\alpha$, a source redshift bin $\beta$ and an angular bin $\theta_k$ is
\begin{equation}
\label{eq:gt_estimator}
    \bar{\boldsymbol{g}}_{\rm t}^{\alpha \beta} (\theta_k) = \frac{\sum_i \sum_j w^{ij} \boldsymbol{e}_t^{ij}(\theta_k)}{\sum_i \sum_j w^{ij} (\mathcal{R}^{ij} + \mathcal{R}_{\rm s}^{\beta})} \bigg|_{i \in \alpha, j \in \beta},
\end{equation}
where $i$ runs over the clusters, $j$ runs over the sources, the weight $w^{ij}$ is the square inverse of the shape measurement error for the $j$-th source galaxy around the $i$-th cluster, $\boldsymbol{e}_{\rm t}^{ij}$ the tangential component of the measured galaxy shape, $\mathcal{R}^{ij}$ is the shear response of the galaxy, and $\mathcal{R}_{\rm s}^{\beta}$ the selection response for the source redshift bin $\beta$ (see, e.g., ref. \cite{Prat2022} for the validation of this lensing estimator). 
We use six logarithmically spaced angular bins ($\theta_k$) over the angular range of $1.1'-20'$ ($0.7'-13'$) for the low (high) redshift bin, which correspond to $0.12-2.19 \, h^{-1} \textrm{Mpc}$ ($0.16-3.06 \, h^{-1} \textrm{Mpc}$) at the minimum redshift of 0.15 (0.43) and $0.26-4.70 \, h^{-1} \textrm{Mpc}$ ($0.21-3.89 \, h^{-1} \textrm{Mpc}$) at the maximum redshift of 0.43 (0.70), respectively. Several factors motivate our choice of angular binning. At small scales, crowdedness of the central regions of clusters can impact the shear measurement through blending.  Additionally, the impact of the brightest cluster galaxy is not included in our mass model, which could bias the lensing inference at small scales. Finally, the boost factors become very large at small scales, making our analysis more sensitive to these non-lensing systematics. In section~\ref{sec:result_systematics}, we test our choice of minimum radial scale by performing an analysis without the first radial bins. We choose the maximum angular scale, on the other hand, to reduce contributions from the two-halo term.  While we do model the two-halo term (see section~\ref{sec:model}), our intent here is to calibrate the cluster masses primarily through the one-halo term. Consequently, our data vectors consist of $6 \,{\rm (angular \, bins)} \times 2 \, {\rm (cluster \, redshift \, bins)} \times 2 \, {\rm (cluster \,} y {\rm \,bins)} \times 2 \, {\rm (source \, redshift \, bins)} = 48$ points.

Since the lensing signal varies with $\Sigma_{\rm crit}$, if information on this quantity for every cluster-galaxy pair is given, we could use this information to weight the tangential shear measurements around the clusters, and thereby improve the signal-to-noise. In a similar vein, one could choose to estimate $\Delta \Sigma$ instead of the tangential shear. Indeed, several studies used this approach \cite[e.g.,][]{Simet2017,McClintock:2019,Robertson2024}. However, we have found that the improvement in signal-to-noise from this alternative approach is small. Moreover, $\Sigma_{\rm crit}$ depends on the redshifts of source galaxies, which are not precisely known. Meanwhile, we account for this uncertainty on the redshift of source galaxies by introducing the redshift shifting parameters in section~\ref{sec:photoz_bias}. Thus, were we to include this weighting to maximize the lensing signal or use $\Delta \Sigma$, our measurements would retain dependence on our model parameters and cosmology. To ensure a clean separation of our measurements from the model, we do not use these approaches. In addition, using bins of projected physical radius requires an assumption of cosmological model during the measurement process to translate angular separation into the projected radius, complicating the subsequent cosmological analysis. By using angular binning, we remove this complication and are able to fully forward model our data vector. The measured lensing data vectors are shown in figure~\ref{fig:gammat_fit} in black.

\subsection{Covariance estimate}
\label{Sec:cov}

\begin{figure}
    \includegraphics[width=0.99\columnwidth]{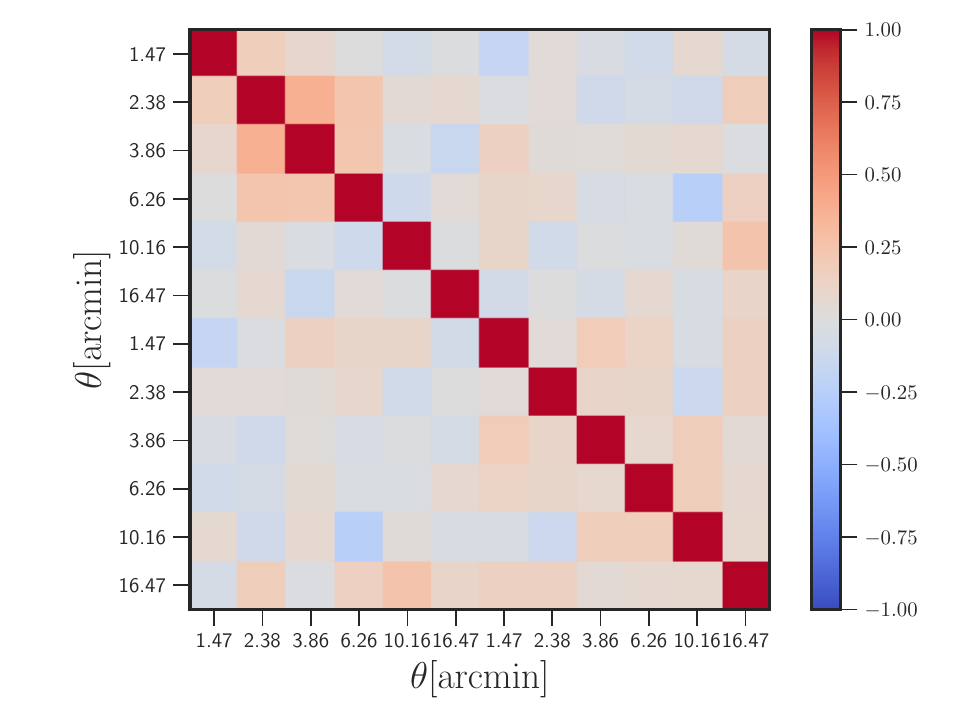}
    \caption{The normalized covariance matrix (correlation matrix) for the tangential shear of the cluster bin $\tilde{y}_0=[0.00,1.05]$ and $z=[0.15,0.43]$. The first six bins represent the tangential shear from the source bin 3, and the last six represents that from the source bin 4.}
    \label{fig:gammat_cov}
\end{figure}

We estimate the covariance matrix of the tangential shear by bootstrapping the clusters with $N$=100,000. All source bins used for each cluster bin are combined when calculating the covariance matrix, so, per dimension, the resultant covariance matrix has the number of source bins (2) times the number of angular bins (6) components (=12). 
We do not include cross-covariance between different cluster bins. Therefore, in total we have four $12\times12$ covariance matrices, corresponding to the four cluster bins in $\tilde{y}_0$ and $z$.
This data-based estimate of the covariance has the advantage that it naturally includes all complexities of our cluster and galaxy selection. This type of data-driven covariance matrix is known to be biased at scales larger than $\sim$30 arcminutes \cite{Friedrich2016}, which is well above our maximum angular scale. 
Note that per cluster bin, the number of our bootstrap samples ($N=100,000$) is significantly larger than the number of data points (12), therefore, a Hartlap-like finite-sample correction \cite{Hartlap2007} for the inverse covariance is negligible (in specific, see section 5 and figure 3 of \cite{Hartlap2007}, which show the bias in inverse covariance from bootstrapping is only $\sim$1\%).

The measured tangential shear data vectors and the corresponding uncertainties for each cluster and source redshift bin combination are shown in the upper panel of figure~\ref{fig:gammat_fit}. Each row represents different bins of Compton-$y$. The left two columns correspond to the low-redshift cluster bin, and the right two columns correspond to the high-redshift cluster bin. For each of the $1\times2$ blocks, the measurement for each source redshift bin (bin 3 on the left and bin 4 on the right) is plotted. The combined SNR across all measurements is 38.8. In addition, as an example, we show the normalized covariance matrix (correlation matrix) for the cluster bin of $\tilde{y}_0 \in [0.00,1.05]$ and $z \in [0.15,0.43]$ in figure~\ref{fig:gammat_cov}. 

\subsection{Lensing null tests}
\label{sec:measurement_null}

\begin{table}
	\centering
	\begin{tabular}{ccc}
    \hline
            Null-$\chi^2$/DoF (PTE) & $z \in$ [0.15,0.43] & $z \in$ [0.43,0.70] \\ \hline
            $\tilde{y}_0 \in$ [0.00,1.05] & 15.9/12 (2.0$\times$$10^{-1}$) & 23.6/12 (2.3$\times$$10^{-2}$) \\
            $\tilde{y}_0 \in$=[1.05,6.00] & 19.3/12 (8.2$\times$$10^{-2}$) & 12.2/12 (4.3$\times$$10^{-1}$) \\
    \hline
	\end{tabular}
    \caption{Null reduced $\chi^2$ values for all the cluster-source bin combinations, for the measurements of the cross shear components (see section~\ref{sec:measurement_null}).}
\label{tab:null_chisq}
\end{table}

\begin{table}
	\centering
	\begin{tabular}{ccc}
    \hline
            Null-$\chi^2$/DoF (PTE) & $z \in$ [0.15,0.43] & $z \in$ [0.43,0.70] \\ \hline
            $\tilde{y}_0 \in$ [0.00,1.05] & 20.1/12 (6.5$\times$$10^{-2}$) & 14.4/12 (2.7$\times$$10^{-1}$) \\
            $\tilde{y}_0 \in$=[1.05,6.00] & 20.0/12 (6.7$\times$$10^{-2}$) & 18.3/12 (1.1$\times$$10^{-1}$) \\
    \hline
	\end{tabular}
    \caption{Null reduced $\chi^2$ values for all the cluster-source bin combinations, for the measurements of the tangential shear around random points (see section~\ref{sec:measurement_null}).}
\label{tab:null_chisq_random}
\end{table}

For any isotropic lens, the cross component of shear, $\gamma_{\rm x} = \gamma_1 \sin 2 \phi - \gamma_2 \cos 2 \phi$, is expected to be zero by angular symmetry. While individual clusters may not be spherically symmetric, our lensing measurements average over many clusters with random orientations.  Consequently, we expect that in the absence of weak-lensing systematics, the measured shear cross component be consistent with null signal. A detection of a significant cross-component of shear would suggest the presence of systematic biases in the lensing measurements.

We show the result for the cross-shear measurements with the grey data points in the upper panel of figure~\ref{fig:gammat_fit} and report the null $\chi^2$ values in table~\ref{tab:null_chisq} along with the probability-to-exceed (PTE) values. While the values of the null-$\chi^2$ are consistent with zero, we note that the cluster bin of $z \in [0.43,0.70]$ and $\tilde{y}_0 \in [0.00,1.05]$ shows a somewhat high $\chi^2$ value with the PTE of 0.023, being only marginally consistent with zero. 

We perform another null test for our shear measurement: tangential shear around random points. Using the same footprint used to select the cluster sample, we draw random points that are 30 times the number of clusters. The result is shown in the light blue data points in figure~\ref{fig:gammat_fit}. We confirm that the results are consistent with zero (see table~\ref{tab:null_chisq_random} for PTE and $\chi^2$ values).

\subsection{Boost Factor}
\label{sec:measurement_boost}

\begin{figure}
    \centering
    \includegraphics[width=0.99\columnwidth]{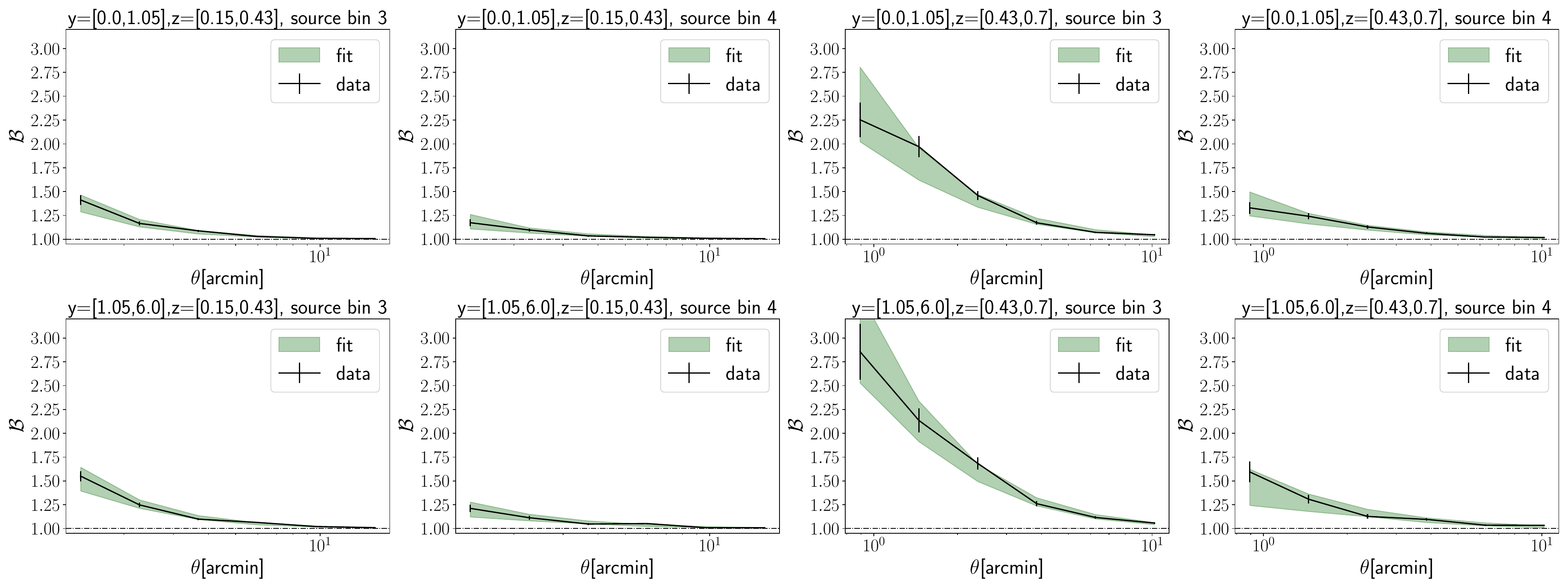}
    \caption{The measured boost factors (black points with errorbars) and the corresponding 68\% confidence intervals (green bands) from the model fitting. The top (bottom) panels correspond to the low (high) Compton-$y$ bins. The left two (right two) columns correspond to the low (high) redshift bins, as indicated in the title.} 
    \label{fig:boost_fit}
\end{figure}

\begin{figure}
    \centering
    \includegraphics[width=0.99\columnwidth]{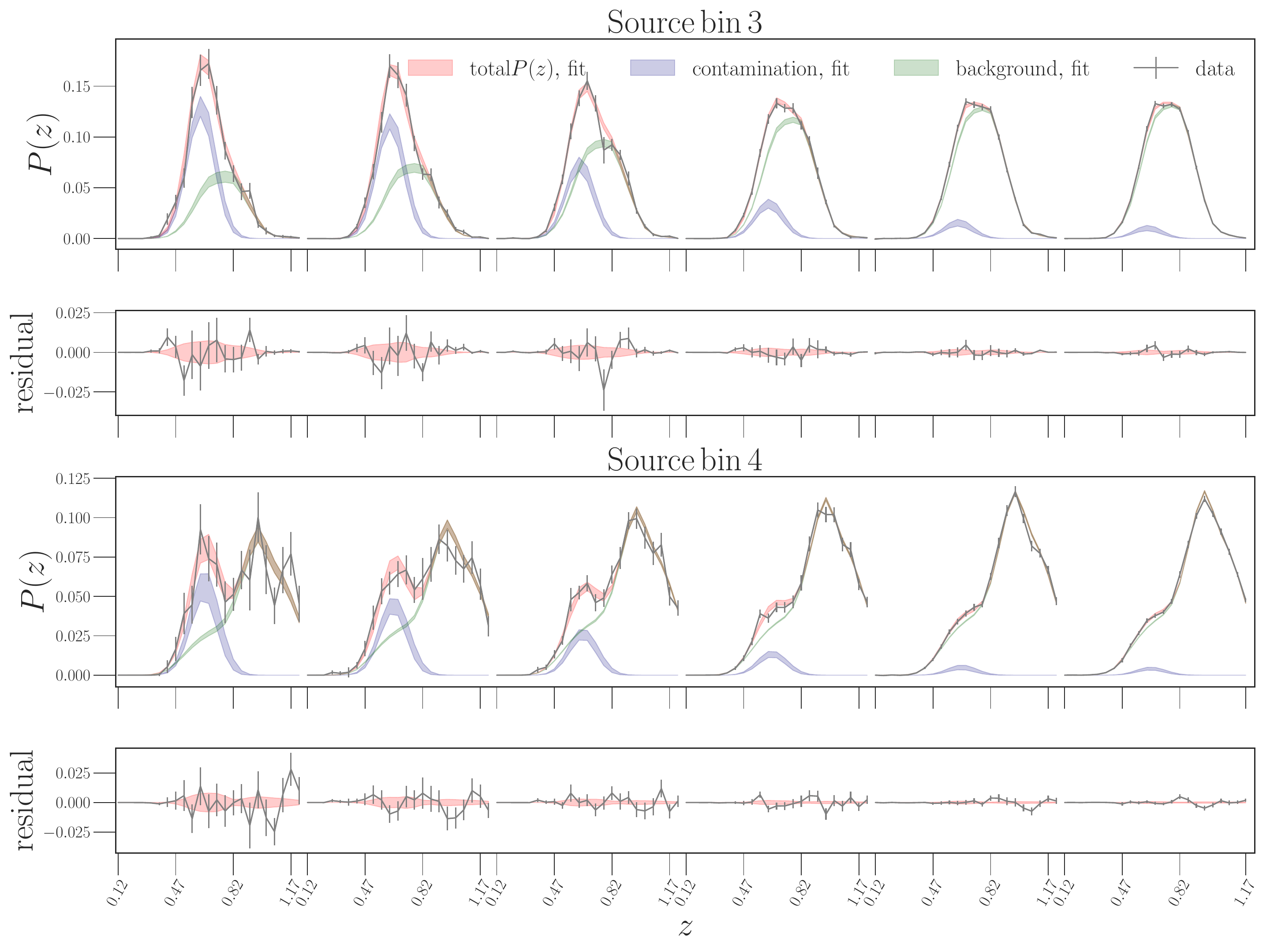}
    \caption{The measured $P(z)$ around the cluster bin of $\tilde{y}_0\in[0.00, 1.05]$ and $z\in[0.43,0.70]$ (grey) and the corresponding fitting result for the $P(z)$ decomposition model for the boost factor (section~\ref{sec:measurement_boost}, red: total, blue: cluster contamination, green: background). Within each panel, different sets of points represent 6 different radial bins, from the smallest to the largest. The top two rows correspond to the measurements with the source bins 3, and the bottom two rows to those of the source bin 4.}
    \label{fig:pz_fit}
\end{figure}

The redshift distribution of source galaxies along lines of sight close to the clusters is likely to deviate from the redshift distribution of all source galaxies. 
This is because the cluster member galaxies and the correlated large-scale structure can contaminate the source galaxy sample due to the photo-$z$ uncertainty. 
Since the cluster members and the foreground galaxies will not be lensed by the clusters, these galaxies act to dilute the lensing signal if included in the source galaxy sample \cite{Gruen2014,Varga2019}. Correcting for this effect (typically a factor of $1-2$ in central regions of clusters, depending heavily on source selection) requires a ``boost factor'' to undo the impact of this dilution.  Properly modeling this effect is crucial for the mass calibration analysis. In this work, we perform a direct measurement of the boost factor, and then use a model fit to these measurements to correct for the boost factor in our lensing analysis.  

We measure the boost factor using $P(z)$ decomposition method \cite{Gruen2014,Varga2019}. We first measure the source redshift distribution, $P(z)$, within each radial bin (see section~\ref{sec:measurement_gammat} for the binning) and fit it with a model that decomposes the measured $P(z)$ into the cluster contamination and the background contribution. We describe further details of the boost factor measurements below, and discuss our model for these measurements in section~\ref{sec:model_boost}.

We measure the boost factor by decomposing the observed distribution of redshift of the source galaxies around our cluster sample into the cluster member and the background components (see ref. \cite{Varga2019} for description and validation of this method). This analysis is done within the same radial bins as our tangential shear estimator, providing the boost factor as a function of cluster-centric angular distance. Specifically, the observed source distribution at a distance of $\theta$, $P(z_{\rm s}|\theta)$ could be expressed as, 
\begin{equation}
\label{eq:boost_pz_model}
    P(z_{\rm s}|\theta) = f_{\rm cl}(\theta) \, P_{\rm cont}(z_{\rm s}|\theta) + (1-f_{\rm cl}(\theta)) \, P_{\rm bg}(z_{\rm s}),
\end{equation}
where $f_{\rm cl}$ is the fraction of the cluster member contamination, $P_{\rm cont}(z_{\rm s}|\theta)$ is the redshift distribution of the cluster members and $P_{\rm bg}(z_{\rm s})$ is the redshift distribution of the background galaxies measured throughout the survey footprint. 
As validated in Varga et al. \cite{Varga2019}, we model $P_{\rm cont}(z_{\rm s}|\theta)$ as a Gaussian distribution with the mean ($\mu_{z, \rm cont}$) and the standard deviation ($\sigma_{z, \rm cont}$) being free parameters. We use the DNF redshift estimates (section~\ref{sec:data_photoz}) to calculate $P(z_{\rm s}|\theta)$ and $P_{\rm bg}(z_{\rm s})$. 
Then the boost factor, $\mathcal{B}$, is related to $f_{\rm cl}$ as,
\begin{equation}
\label{eq:fcl_to_B}
    \mathcal{B} = \frac{1}{1-f_{\rm cl}},
\end{equation}
which relates the observed tangential shear ($\boldsymbol{g}_{\rm obs}$) to the unbiased tangential shear ($\boldsymbol{g}_{\rm true}$):
\begin{equation}
    \boldsymbol{g}_{\rm obs} = \frac{\boldsymbol{g}_{\rm true}}{\mathcal{B}}.
\end{equation}
Note that since $f_{\rm cl}$ is a fraction which is not related to the exact location of the member galaxies in the photometric redshift space, our calculation of the boost factor is not affected by the bias in the DNF redshift estimates 
(see section 3 and the corresponding figures of \cite{Varga2019} for the full validation of this method; also see section 3.3.1 of \cite{Grandis2024lensing} and the references therein).\footnote{Varga et al. \cite{Varga2019} used BPZ \cite{BPZ} to validate the method. However, given the smaller photo-z bias and scatter of DNF than those of BPZ for the DES-Y3 galaxies \cite{SevillaNoarbe2021}, it provides enough validation for our DNF case.}

Given a set of source galaxies, $P(z_{{\rm s},j})$ for the $j$-th redshift bin is calculated as,
\begin{equation}
    P(z_{{\rm s},j}) = \frac{\sum_i w^{ij} (\mathcal{R}^{ij} + \mathcal{R}_{\rm s}) }{\Delta z_{\rm s} \sum_i \sum_k w^{ik} (\mathcal{R}^{ik} + \mathcal{R}_{\rm s})},
\end{equation}
where $i$ runs over the source galaxies in the $j$-th redshift bin and $\Delta z_{\rm s}$ is the width of the redshift bins, which we set to 0.05. The $k$ runs over all redshift bins between $z$=0 and 3.\footnote{Note that the ``redshift bins'' here represent the binning of $P(z_{\rm s})$, not the four source redshift bins defined in section~\ref{sec:data_photoz}}

To fit the measured redshift distribution to the model in eq.~\ref{eq:boost_pz_model} to obtain the boost factor data vector, we calculate $P(z_{\rm s}|\theta)$ and $P_{\rm bg}(z_{\rm s})$ over the redshift range between 0.10 and 1.25 with $dz=0.05$. The corresponding uncertainties on $P(z_{\rm s}|\theta)$ are calculated by bootstrapping the cluster sample with $N$=100,000. Note that the uncertainty on $P_{\rm bg}(z_{\rm s})$ is negligible with respect to that of $P(z_{\rm s}|\theta)$ due to the significantly higher number of source galaxies. We therefore ignore the uncertainty on $P_{\rm bg}(z_{\rm s})$. 

Using this estimated uncertainty on $P(z_{\rm s}|R)$, we find the best-fit values for $f_{\rm cl}$, $\mu_{z, \rm cont}$ and $\sigma_{z, \rm cont}$ for each bootstrapped set of \{$P(z_{\rm s}|\theta)$, $P_{\rm bg}(z_{\rm s})\}$, from which we calculate the final covariance matrix for the boost factor ($\mathcal{B}$) using eq.~\ref{eq:fcl_to_B}. 
Note that, per cluster bin, $\mu_{z, \rm cont}$ and $\sigma_{z, \rm cont}$ are shared across all radial bins, but we allow different sets of \{$\mu_{z, \rm cont}$, $\sigma_{z, \rm cont}$\} for different source redshift bins. Since we have 6 radial bins and 2 source redshift bins, there are 16 (6$\times$2 + 2$\times$2) free parameters per cluster bin for this calculation.

\begin{figure}
    \includegraphics[width=0.99\columnwidth]{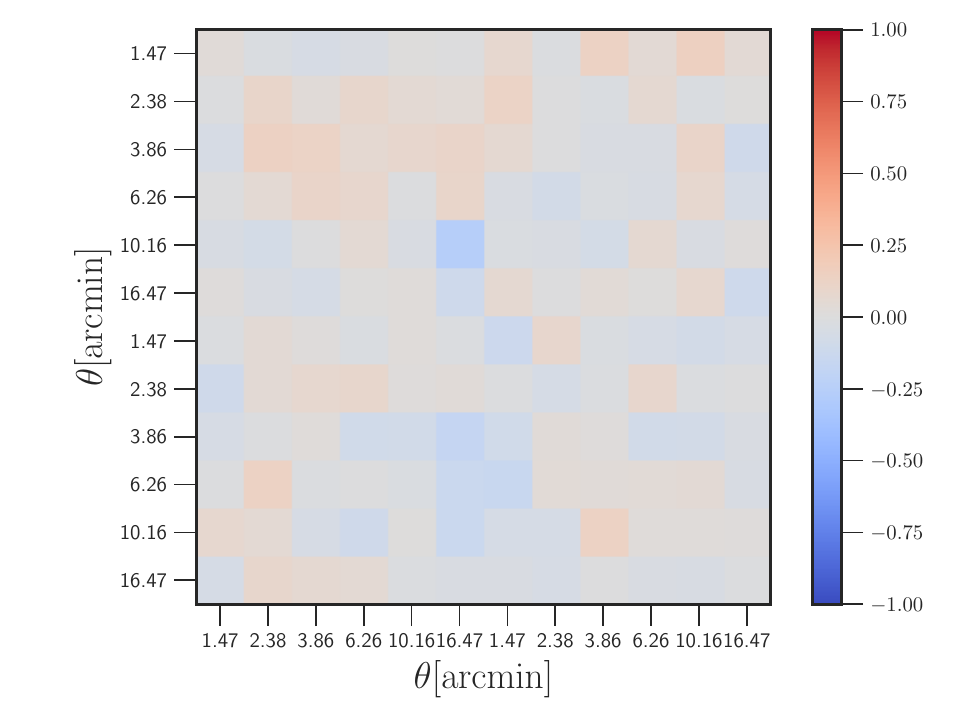}
    \caption{The normalized covariance matrix (correlation matrix) \textit{between} the boost factor (x-axis) and the tangential shear (y-axis) of the cluster bin $\tilde{y}_0=[0.00,1.05]$ and $z=[0.15,0.43]$. The first six bins correspond to the source bin 3, and the last six bins to the source bin 4. We do not find any significant cross components.}
    \label{fig:boost_cov}
\end{figure}

The resultant boost factors are shown in figure~\ref{fig:boost_fit}.
Each row represents different Compton-$y$ bins, while the left two columns correspond to the low-redshift cluster bin and the right two columns to the high-redshift cluster bin. As expected, the boost factors asymptote to 1 (no cluster member contamination) at large scales, increasing as approaching to the cluster center. 
In addition, as a sanity check, in figure~\ref{fig:pz_fit}, we plot the measured $P(z_{\rm s}|\theta)$ around the cluster bin of $\tilde{y}_0\in[0.0,1.05]$ and $z\in[0.43,0.7]$ as an example, along with the best-fit model components (total in red, cluster contamination in blue, and background contribution in green), which demonstrates a reasonable fitting. Note that we plot all six radial bins in one plot.

Furthermore, in figure~\ref{fig:boost_cov}, we show the normalized cross-covariance matrix (correlation matrix) \textit{between} the tangential shear and the boost factor for the cluster bin of $\tilde{y}_0 \in [0.0, 1.05]$ and $z \in [0.15, 0.43]$. For all cluster bins, we do not find any significant correlation between the tangential shear and the boost factor. Therefore, we treat the tangential shear and the boost factor as effectively uncorrelated data vectors from each other in the following analysis.

\section{Modeling}
\label{sec:model}

In this section, we describe how we construct a model for the tangential shear and the boost factor measurements. 

\subsection{Model for 3D mass distribution}
\label{sec:halo_model}

Our model for the tangential shear begins with a model for the cluster mass distribution. We assume that the mass distribution (when averaged across the clusters in a bin) is close to being spherically symmetric so that it could be written as $\rho(r)$, where $r$ is the separation from the cluster center in 3D space.

We separate $\rho(r)$ into the so-called one-halo and two-halo terms:
\begin{eqnarray}
    \rho(r) = \rho^{1h}(r) + \rho^{2h}(r).
\end{eqnarray}
Here, the one-halo term denotes the contribution from the halo of interest itself, while the two-halo term represents the contribution from the neighboring halos and the large-scale structures. 

Baryons constitute approximately a sixth of the total mass of galaxy clusters. Their distribution is generally different from that of dark matter. Therefore, not accounting for baryons in a weak-lensing mass calibration analysis could lead to a biased result \cite{Henson2017,Lee2018}. We follow the prescription proposed by Cromer et al. \cite{Cromer2022} to separate the contribution of baryons from the total matter density:
\begin{eqnarray}
    \rho^{1h}(r) = \rho^{\rm b}(r) + \rho^{\rm CDM}(r),
\end{eqnarray}
where
\begin{eqnarray}
    \rho^{\rm CDM}(r) = \frac{(1-f_{\rm b})\rho_{\rm 0}}{(r/r_{\rm s)}(1+r/r_{\rm s})^2}
\end{eqnarray}
and
\begin{eqnarray}
\label{eq:baryon}
    \rho^{\rm b} = \frac{f_{\rm b} \rho_{\rm b}}{(r/r_{\rm s}x_{\rm c})^\gamma [1+(r/r_{\rm s}x_{\rm c})^{1/\alpha}]^{(\beta-\gamma)\alpha}}.
\end{eqnarray}
Note that $\rho^{\rm CDM}$ is an NFW profile \cite{Navarro1997} with a modification by a factor of $1-f_{\rm b}$ (the dark matter fraction), and $\rho^{\rm b}$ is a generalized NFW (GNFW) profile \cite{Zhao1996} with a modification factor of $f_{\rm b}$. 
Here, $f_{\rm b}=\Omega_{\rm b}/\Omega_{\rm m}$ is the cosmic baryon fraction, $\rho_{\rm 0}$ is the normalization factor for the NFW profile, and $r_{\rm s}=r_{\rm 500c}/c_{\rm 500c}$ is the scale radius, where $c_{\rm 500c}$ represents the halo concentration using $R_{\rm 500c}$ definition of the halo boundary. 
Also, $\rho_{\rm b}$ is the normalization factor for the GNFW profile, $x_{\rm c}$ is the core scale fixed at 0.5, and $\{\gamma, \alpha, \beta\}$ are the cluster core power-law, transition power-law and long-distance power-law, respectively, where $\gamma$ is fixed at 0.2.\footnote{Note that the NFW scale radius for a halo of our mean mass, $M_{\rm 500c} = 5.4 \times 10^{14} M_{\odot}$, with $c_{\rm 500c}=2.45$ at $z=0.4$ is $\sim$$0.3 h^{-1}{\rm Mpc}$. Therefore, the effective influence range of $\gamma$ is $r \ll r_{\rm s}x_{\rm c} \sim 0.15 h^{-1}{\rm Mpc}$, which lies below our minimum measurement scale (section~\ref{sec:measurement_gammat}). Hence, $\gamma$ does not meaningfully change our model prediction so that we fix it at 0.2.}
We fix the concentration of the NFW profile at 2.45 according to the concentration-mass relation from Diemer \& Joyce \cite{Diemer2019}. We have checked that the choice of concentration does not significantly affect our result. Note that this model is similar but not identical to the widely used ``baryonification'' models \cite{Schneider2019,Giri2021,To2024,Anbajagane2024}.

The two-halo term is modeled as follows:
\begin{eqnarray}
    \rho^{2h}(r) = a_{2h} \rho_{\rm m} b(M) \xi_{\rm mm}(r,z).
\end{eqnarray}
Here, $\rho_{\rm m}$ is the mean matter density of the universe, $b(M)$ is the halo bias modeled in Tinker et al. \cite{Tinker2010}, and $\xi_{\rm mm}(r,z)$ is the matter-matter correlation function of the universe. Note that our model includes the factor of $a_{2h}$ to account for any deviation of the halo bias, $b(M)$, from the theoretical prediction; $a_{2h}=1$ corresponds to the case where the halo bias is the same as the value calculated in Tinker et al. \cite{Tinker2010}.

This halo model with baryons has been validated to be unbiased at $\lesssim 1\%$ level on the WL mass calibration of galaxy clusters, as shown in Cromer et al. \cite{Cromer2022}, where they used 1,400 simulated galaxy clusters from the hydrodynamical simulation suites from ref. \cite{Battaglia2010} with a very similar treatment of modeling stacked weak-lensing signal as ours (modeling individual profiles and stacking), although they have not included the effect of miscentering.
We refer the reader to Cromer et al. \cite{Cromer2022} for further details of this model as well as its implementation. This model is implemented as a \texttt{Python} package \textsc{maszcal}.

\subsection{Modeling the lensing observables}
\label{sec:model_gt}
Under the Born approximation\footnote{Projection of the 3D density along unperturbed photon paths, neglecting multiple deflections.}, which should be valid for the lensing measurements considered here, the lensing signal is related to the line of sight projection of the mass distribution, $\Sigma(R)$, given by
\begin{equation}
\Sigma(R) = \int_{-\infty}^{\infty}\rho(\sqrt{R^2 + l^2})\,dl,
\end{equation}
where $R$ is the projected distance from the cluster center and $\Sigma(R)$ is related to the reduced tangential shear ($\boldsymbol{g}_{\rm t}$) defined in section~\ref{sec:measurement_gammat}.

In reality, the lens (cluster) and source galaxy redshifts are not known exactly. Rather, we characterize these with PDFs $P(z_{\rm l})$ and $P(z_{\rm s})$, respectively. We then integrate over these distributions to obtain the expected $\boldsymbol{g}_t$.  

For the $i$-th cluster with the $j$-th source redshift bin at the cluster-centric radial angle of $\theta$,
\begin{equation}
\label{eq:gammat_model_ind}
\hat{\boldsymbol{g}}_{t,ij} (\theta) =
\int dz_s \int dz_l \, n_{l,i} (z_l) n_{s,j}(z_s;\vec{\theta}_{z}) \frac{\Delta \Sigma(\theta;\vec{\theta}_M)}{\Sigma_{\rm crit}(z_l, z_s)-\Sigma(\theta)} \bigg\rvert_{z_s > z_l},
\end{equation}
where $n_{l,i}(z)$\footnote{For the lenses, we assume a Gaussian distribution of the cluster redshift, as reported in the catalog. However, note that since the redshift uncertainty of the clusters are much smaller than the width of the source redshift distribution, the effect of including cluster redshift uncertainty is minimal, which we have checked.} and $n_{s,j}(z;\vec{\theta}_{z})$ are the cluster and the source galaxy redshift distributions, respectively, which depend on the redshift systematics parameters $\vec{\theta_{z}}$, and $\vec{\theta}_M$ denotes the parameters describing the halo model (section~\ref{sec:halo_model}), $\{ M, \alpha, \beta, a_{2h} \}$ and $\{ f_{\rm mis}, R_{\rm mis} \}$ (see section~\ref{sec:miscentering} for the last two parameters related to the halo miscentering).

The final model of $\boldsymbol{g}_t$ for a given bin of clusters and the $j$-th source redshift bin is then
\begin{eqnarray}\label{eq:gt_mean_sys}
\hat{\boldsymbol{g}}_{t,j} = [(1+m_{\gamma,j}) / \mathcal{B}_j] \sum_i w_i \hat{\boldsymbol{g}}_{t,ij}
\end{eqnarray}
where $m_{\gamma,j}$ is the multiplicative shear bias, $\mathcal{B}_j$ the boost factor from the $j$-th source redshift bin, and $w_i$ the sum of the weights of the source galaxies (the summand of the denominator of the eq.~\ref{eq:gt_estimator}) for the $i$-th cluster, normalized to unity in summation.
Note that we also forward-model the boost factor instead of correcting the tangential shear measurement.

\subsection{Relating the halo mass to the SZ observable: hydrostatic bias}
\label{sec:hs_bias}

We model the hydrostatic mass bias, $1-b = M_{\rm SZ}/M_{\rm true}$ as follows. 
The true mass, $M_{\rm true}$, is related to the mass from the SZ scaling relation, $M_{\rm SZ}$, as,
\begin{equation}
\label{Eq:Mtrue}
     \frac{M_{\rm true}}{M_{\rm SZ}} = \frac{1}{1-b} = A_{\rm mass} \left(\frac{M_{\rm SZ}}{3 \times 10^{14} M_{\odot}} \right)^{\eta} \left(\frac{1+z}{1+0.45} \right)^{\zeta},
\end{equation}
where $A_{\rm mass}$, $\eta$ and $\zeta$ are free parameters. We also consider the case where the hydrostatic mass bias does not carry any mass or redshift dependence, represented by one number over the whole redshift and mass range, $A_{\rm mass}=1/(1-b)$. Note that this power-law modeling of the hydrostatic mass bias has been explored before, e.g., by refs. \cite{Salvati2019,Pandey2022,Wicker2023}, although with different methodologies, as well as that $1-b \propto M_{\rm SZ}^{-\eta} (1+z)^{-\zeta}$.

For the values of $M_{\rm SZ}$ of each cluster, we take the SZ scaling relation, $P(M_{\rm SZ}|\tilde{y}_0)$, of the ACT DR5 clusters reported by Hilton et al. \cite{Hilton2021}. We fix the scatter in $\ln (M_{\rm SZ})$ given $\tilde{y}_0$ at $\sigma_{\ln M}=0.2$, assuming a log-normal distribution of the mass around the mean value. We have checked that setting the mass scatter free over a reasonable range does not affect our posterior for the hydrostatic mass bias (see section~\ref{sec:result}).

The figure 7 of Hilton et al.~\cite{Hilton2021} shows that the completeness of our cluster sample is nearly a flat function of redshift. Therefore, we expect the impact of redshift-dependent completeness on the inferred value for $\zeta$ to be minimal.

However, note that while the cluster detection, characterization and calibration of $M_{\rm SZ}$ in Hilton et al.~\cite{Hilton2021} accounted for a wide range of systematics -- such as the CMB noise properties, survey depth inhomogeneity, angular scale mismatch between the filter and cluster size, relativistic corrections, intrinsic scatter, and Eddington bias correction -- the adoption of the universal pressure profile (UPP, \cite{Arnaud2010}) could in principle introduce redshift-dependent selection effect if the deviation of the cluster gas profile from the UPP evolves with redshift. In addition, although the projection from the halos along the line of sight is suppressed in the SZ signal ($y\sim M^{5/3}$), unmodeled projection effects could still influence the halo selection possibly in a redshift-dependent manner. Furthermore, Hilton et al. \cite{Hilton2021} assumes a constant intrinsic scatter of 0.2, which we allow to vary. However, if the intrinsic scatter evolves with redshift, the associated Eddington bias correction would need to be updated. Finally, cluster concentration and triaxiality may induce additional scatter to the SZ scaling relation \cite{Baxter2024,Gallo2024,Saxena2025}, although their impact on cluster selection and scaling relation bias remains an active area of ongoing research. We note these possible systematics for completeness, but we defer detailed investigations of these systematics to future cosmological studies.

\subsection{Modeling the boost factor}
\label{sec:model_boost}

In order to avoid multiplying the data vectors by noisy measurements and to fully forward-model the WL observable, we simultaneously fit the boost factor measurements and the WL measurements by adopting a functional form for the boost factor, as described below. Note that in this analysis, the boost factors are applied to the model, not to the data vectors.

To model the boost factor measurements described in section~\ref{sec:measurement_boost} we use the model from McClintock et al. \cite{McClintock:2019}, which is equivalent to an NFW profile \cite{Navarro1996}:
\begin{eqnarray}
    \mathcal{B}(x) = 1 + B_0 \frac{1-F(x)}{x^2 - 1},
\end{eqnarray}
with 
\begin{eqnarray}
    x = R/R_s
\end{eqnarray}
and 
\begin{eqnarray}
    F(x) = \begin{cases}
        \frac{\tan^{-1}\sqrt{x^2-1}}{\sqrt{x^2-1}} \, & (x>1) \\
        1 \, & (x=1) \\
        \frac{\tanh^{-1}\sqrt{x^2-1}}{\sqrt{1-x^2}} \, & (x<1)
    \end{cases}
\end{eqnarray}
where $R_s$ and $B_0$ are free parameters of the model. The boost factor model is then applied to the model for $\boldsymbol{g}_t$ (see section~\ref{sec:model_gt}, eq.~\ref{eq:gt_mean_sys}).

\subsection{Systematic uncertainties}
\label{sec:sys_uncertainty}

Several sources of systematic uncertainty contribute to our analysis. These include biases to the photo-z distributions of the source galaxies, biases in the WL shear calibration, and miscentering of the SZ-selected clusters. We parameterize these systematic uncertainties into our model. We discuss each of these sources of systematic uncertainty in more detail below.

\subsubsection{Photo-$z$ bias}
\label{sec:photoz_bias}

Estimating the distribution of the source galaxy redshift (section~\ref{sec:data_photoz}) relies on photometry, and therefore is subject to potential biases. Abbott et al. \cite{DES3x2} showed that the photo-$z$ bias simply modeled with a shift parameter $\Delta z_s$ produces unbiased results in 3$\times$2-point cosmology. Since the cluster lensing shares a similar theoretical framework as the galaxy-galaxy lensing in the 3$\times$2-point analysis, we assume that we could use this treatment to the cluster lensing in this analysis.\footnote{Note that galaxy clusters reside in much more non-linear environments than the galaxies in a typical galaxy-galaxy lensing analysis, which could potentially make cluster lensing sensitive to the shape of $n(z)$. Therefore, further validity of this assumption requires more rigorous validation tests, which we defer to a future study.} That is, 
\begin{equation}
    n_s(z_s;\vec{\theta}_{z}) = n_s(z_s - \Delta z_s),
\end{equation}
for each source redshift bin (see eq.~\ref{eq:gammat_model_ind} in section~\ref{sec:model_gt}), where $n_s(z_s)$ represents the measured distribution of the source galaxy redshift (section~\ref{sec:data_photoz}). We follow this treatment in our model, adopting the values and uncertainties for $\Delta z_s$ from ref. \cite{DES3x2}, from which we randomly sample values for $\Delta z_s$ per model calculation when running MCMC chains for the main analysis. \footnote{Note that we don't include $\Delta z_s$ as a free parameter in the model, but use values randomly sampled from the previous DES cosmology analysis, which are rigorously tested with simulations and mock data.} The mean and the standard deviation for $\Delta z_s$ we adopt is 0 (0) and 0.011 (0.017) for the $3^{\rm rd}$ ($4^{\rm th}$) source redshift bin, respectively.

\subsubsection{Shear calibration bias}
\label{sec:shear_bias}

The shape measurements for the source galaxies are also subject to potential biases especially due to blending which could bias the measured galaxy ellipticity and impact the source selection \cite[e.g.,][]{Samuroff2018,Sheldon2020,MacCrann2022,Yamamoto2025,Zhang2025shear}. As described in section~\ref{sec:data_shear}, calibration on these multiplicative shear biases ($m_\gamma$) has been calculated for the DES-Y3 source galaxy selections \cite{DES3x2,MacCrann2022}, which we also use for our analysis. Therefore, we use the same distributions of multiplicative shear bias parameters by randomly sampling values from the distributions reported in refs. \cite{MacCrann2022,DES3x2} when running MCMC chains. The mean and standard deviation for $m_\gamma$ we adopt is -0.024 (-0.037) and 0.008 (0.008) for the $3^{\rm rd}$ ($4^{\rm th}$) source redshift bin, respectively. 

Note that for the photo-$z$ and the shear calibration bias parameters, cluster weak-lensing exhibits negligible level of constraining power due to their degeneracy to the cluster mass itself. Therefore, even if we have explicitly included those parameters in our chain, our posterior contours would be almost identical to the priors (see, e.g., the parameter $\mathcal{A}_m$ in section 5.3.3 and figure 10 of McClintock et al. \cite{McClintock:2019}). Therefore, we choose not to explicitly include these bias parameters in our MCMC chains.

\subsubsection{Miscentering}
\label{sec:miscentering}

We use the positions of the brightest central galaxies (BCGs) as the centers of our cluster sample for the WL measurement. However, BCGs could be misidentified or deviate from the true cluster centers \cite[e.g.,][]{Zhang2019miscentering,Kelly2023,Sommer2024}. Therefore, in order to obtain an unbiased result on the mass calibration, one must include this miscentering effect in the halo model \cite[e.g.,][]{McClintock:2019,Grandis2024lensing,Bocquet2023}. Since most of these BCG confirmations come from the galaxy clusters identified by the Dark Energy Survey \cite{Hilton2021}, we use the same miscentering model as adopted in McClintock et al. \cite{McClintock:2019}.

For a sample of clusters, the stacked surface mass density $\Sigma$ can be expressed as,
\begin{eqnarray}
\Sigma = \frac{\sum_i \left[ w_i (1 - f_i) \Sigma_i + w_i f_i \Sigma^{\rm mis}_i \right]}{\sum_i w_i},
\end{eqnarray}
where $f_i = 0$ if the cluster is centered, and $f_i = 1$ if the cluster is miscentered. $\Sigma_i$ and $\Sigma_i^{\rm mis}$ are the surface mass density profile of the $i$-th cluster and that when a miscentered position is used as the center, respectively. Here, $w_i$ is the weight applied to the data measurement. Note that $\Sigma^{\rm mis}$ depends on how much the cluster is miscentered. Then the expectation value for this quantity over a given miscentering distribution is 
\begin{eqnarray}
\langle \Sigma \rangle &=& \frac{\sum_i \left[ w_i \langle (1 - f_i) \Sigma_i \rangle + w_i \langle f_i \Sigma^{\rm mis}_i \rangle \right]}{\sum_i w_i} \\
&=& (1 - f_{\rm mis}) \frac{\sum_i w_i\Sigma_i}{\sum_i w_i} + \frac{\sum_i w_i \langle f_i \Sigma_i^{\rm mis} \rangle}{\sum_i w_i}
\end{eqnarray}
where $f_{\rm mis}$ is the miscentered fraction, i.e. $f_{\rm mis} \equiv \langle f_i \rangle$, the expectation value of $f_i$, which we will use to generate our model as explained below.
We then have
\begin{eqnarray}
\langle \Sigma \rangle &=&  (1 - f_{\rm mis}) \frac{\sum_i w_i\Sigma_i}{\sum_i w_i} + f_{\rm mis} \frac{\sum_i w_i \langle \Sigma_i^{\rm mis} \rangle}{\sum_i w_i},
\end{eqnarray}
where $\Sigma^{\rm mis}$ is expressed as,
\begin{equation}
\label{eq:Sigma_RRmis}
\Sigma^{\rm mis}(R \, | \, R_{\text{mis}}) = \frac{1}{2 \pi}\int_{0}^{2 \pi} {{\rm d}\phi \, \Sigma\left(\sqrt{R^2 + R_{\text{mis}}^2 + 2 R R_{\text{mis}} \cos{\phi}}\,\right)}.
\end{equation}
Here, $R_{\rm mis}$ is the miscentering amplitude and $\phi$ is the azimuthal angle. Note that $w_i$ here is the same weight as in eq.~\ref{eq:gt_mean_sys}.

Following McClintock et al. \cite{McClintock:2019} (also see ref. \cite{Zhang2019miscentering}), we assume a Gamma distribution for $R_{\rm mis}$:
\begin{equation}
    P(R_{\rm mis})= \frac{R_{\rm mis}}{\sigma_{\rm mis}^2} \exp{\Big( -\frac{R_{\rm mis}}{\sigma_{\rm mis}}\Big)}.
\end{equation}

For a given set of miscentering parameters, $f_{\rm mis}$ and $\sigma_{\rm mis}$, we statistically realize this miscentered profile as follows. For each cluster in the model, we determine whether the cluster is miscentered or not according to the miscentering probability $f_{\rm mis}$. If the cluster is determined to be miscentered, we then draw a random number for $R_{\rm mis}$ according to the probability distribution above to calculate the miscentered density profile as in eq.~\ref{eq:Sigma_RRmis}. We finally calculate the stacked tangential shear model with these cluster density profiles (section~\ref{sec:model_gt}), while repeating these random draws per cluster per model calculation when running MCMC chains. 

Note that one could potentially reduce the fraction of miscentered clusters by selecting clusters whose BCG positions agree with those of the tSZ signal (therefore less turbulence and mergers; ref. \cite{Ding2025}). However, doing so will alter the cluster number count, which is difficult to model and involves significant reduction in the WL SNR. While this approach is astrophysically interesting, we do not pursue such further selection in this cosmological study.

\section{Hierarchical Bayesian model}
\label{sec:likelihood}

\subsection{Likelihood}
\label{sec:like_detail}

We now perform a Bayesian analysis of the tangential shear and boost factor measurements in order to constrain the parameters of our model, namely the hydrostatic bias parameters. We note that the model for the tangential shear depends on the boost factor, but that the boost factor and tangential shear measurements are close to independent. 
Even if the likelihood of the tangential shear has additional constraining power on the boost factor parameters, it would not meaningfully alter the mass constraint since they are not degenerate (see figure 10 of \cite{McClintock:2019} which shows nearly independent posteriors between the cluster mass and the boost factor parameters).

We treat the measurements for each cluster redshift bin as independent.  This is reasonable since the redshifts of the clusters are constrained to significantly higher precision than the bin widths, and the correlation functions of interest are small at the scales corresponding to the bin width.  Below, the likelihood $P$ should be understood to correspond to a single cluster redshift bin (although we will also present results that combine constraints across all redshift bins).  Note that we include cross-covariance between source galaxy redshift bins and between angular bins of the tangential shear and boost factor measurements.

Treating the boost factor and the lensing measurements as independent, we write the total log-likelihood for both as the sum of their individual log-likelihoods:
\begin{equation}
\label{eq:joint_like}
   \ln \mathcal{P}(\hat{\boldsymbol{g}}_t, \hat{\mathcal{B}} |\theta_{\gamma}, \theta_{\rm boost}, \{Y_{\rm obs}\}) = \ln \mathcal{P}(\hat{\boldsymbol{g}}_t|\theta_{\gamma},\theta_{\rm boost}, \{Y_{\rm obs}\}) +  \ln \mathcal{P} (\hat{\mathcal{B}}|\theta_{\rm boost}),
\end{equation}
where $\gamma$ and `boost' subscripts indicate the lensing and boost factor, respectively, the hats denote measured quantities, and the $\theta$ denotes sets of parameter values.  For the boost factor, the parameters are clear: $\theta_{\rm boost} = \{R_s, B_0\}$ (section~\ref{sec:model_boost}) and we assume a Gaussian likelihood around the expected model. Here, $Y_{\rm obs}$ includes $\tilde{y}_{0,i}$ and the redshift of the cluster, $Y_{{\rm obs},i} = \{\tilde{y}_{0,i}, z_{l,i}\}$.

The choice of parameters for the lensing measurements is more subtle.  Our model for the lensing signal is written in terms of the cluster mass, $M$. 
However, each cluster bin contains multiple clusters covering a certain distribution of masses.  Moreover, we ultimately wish to constrain the hydrostatic bias parameter, not just the cluster masses.  Furthermore, the relation between the SZ signal and $M$ includes scatter, which must be included in our analysis.  

We take a hierarchical approach to this problem. The mass of each cluster is drawn from a distribution $ P(M_i | \theta, Y_{{\rm obs},i})$, where $Y_{\rm obs}$ here represents both the cluster $\tilde{y}_0$ value and the redshift. This distribution encapsulates the mass-observable relationship, and we will parameterize it as a log-normal distribution:
\begin{equation}
\label{eq:P_M_Y}
P(\ln M_i | \theta, Y_{{\rm obs},i}) \propto \exp \left[-\frac{(\ln M_i - \ln \bar{M}_{\rm true}(\theta))^2 }{2\sigma_{\ln M}^2} \right],
\end{equation}
where $\bar{M}_{\rm  true}$ is given by eq.~\ref{Eq:Mtrue} and the scatter, $\sigma_{\ln M}$, is set by the scatter in the SZ scaling relation, $P(M_{\rm SZ}|Y_{\rm obs})$, which is fixed at 0.2\footnote{Note that, in section~\ref{sec:result_systematics}, we check that freeing up this scatter parameter does not meaningfully affect our result.}. 
Note that, as briefly described in section~\ref{sec:hs_bias}, we do not explicitly model the scaling relation $P(M_{\rm SZ}|Y_{\rm obs})$ but rather take the reported $M_{\rm SZ}$ values in the catalog \cite{Hilton2021} while adding the scatter to the values. Therefore, while we include $P(M_{\rm SZ}|Y_{\rm obs})$ in the following equations for completeness, the calculation of this scaling relation is implicitly done by Hilton et al. \cite{Hilton2021}, which we do not try to modify, but to which we calculate the correction factor ($A_{\rm mass}$, section~\ref{sec:hs_bias}). 

The observed $\hat{\boldsymbol{g}}_t$ is assumed to be normally distributed around the expectation value given a set of individual cluster masses, $\{M_i\}$, and the free parameters, $\theta$, with the covariance given in section~\ref{Sec:cov}:
\begin{multline}
    P(\hat{\boldsymbol{g}}_t | \langle \boldsymbol{g}_t \rangle_{\{M_i\},\theta}) = P(\hat{\boldsymbol{g}}_t | \{M_i(M_{{\rm SZ},i}, \theta)\} ) \propto \\
    \exp \left[-\frac{1}{2}(\hat{\boldsymbol{g}}_t - \langle \boldsymbol{g}_t \rangle_{\{M_i\}, \theta})^T \Sigma^{-1} (\hat{\boldsymbol{g}}_t - \langle \boldsymbol{g}_t \rangle_{\{M_i\}, \theta})  \right],
\end{multline}
where $\theta$ includes all the free parameters in this analysis (mass correction, 2-halo amplitude, miscentering, baryonic parameters and boost factor parameters).
Here, we use $\boldsymbol{g}_t$ to refer to the reduced tangential shear profile.
The expectation value of $\boldsymbol{g}_t$ for each cluster bin is given by taking a weighted average over all clusters in the bin as described in section~\ref{sec:model_gt}.

Ultimately, our main interest is constraining the parameters of the mass-observable relation, rather than the individual cluster masses.  We thus marginalize over the individual mass parameters, $M_i$:
\begin{equation}
    \mathcal{P}(\theta | \hat{\boldsymbol{g}}_t, \{Y_{\rm obs}\}) \propto \mathcal{P}(\theta) \mathcal{P}( \hat{\boldsymbol{g}}_t | \theta, \{Y_{\rm obs}\} ), 
\end{equation}
where
\begin{equation}
\label{eq:likelihood}
    \mathcal{P}( \hat{\boldsymbol{g}}_t | \theta, \{Y_{\rm obs}\} ) = \int dM_1 \ldots dM_N  P( \hat{\boldsymbol{g}}_t | \{M_i(M_{SZ,i},\theta)\}) P(\{M_{SZ,i}\} | \{Y_{\rm obs}\}),
\end{equation}
where $P(\{M_{SZ,i}\} | \{Y_{\rm obs}\})$ is the observable-mass relation given in Hilton et al. \cite{Hilton2021}, and $P(\hat{\boldsymbol{g}}_t | \{M_i(M_{SZ,i},\theta)\})$ is given by our halo model and the model for the hydrostatic mass bias (section~\ref{sec:model}).
This is an $N_{\rm cluster}$-dimensional integral, which involves a considerable amount of computing time. We find that the integral can be approximated with a Monte Carlo approach in a stochastic manner. That is,
\begin{eqnarray}
 \label{eq:likelihood_approx}
 \mathcal{P}(\hat{\boldsymbol{g}}_t | \theta, \{Y_{\rm obs}\}) &\approx& \frac{\sum_j^{N_{\rm draws}} P(\hat{\boldsymbol{g}}_t | \{M_i (M_{SZ,i},\theta)\} )  }{N_{\rm draws}}
\end{eqnarray}
where we use $N_{\rm draws}=7$ and each draw is a random realization of $\{ M_i \}$ from $P(M_{SZ} | Y_{\rm obs})$ with a mass correction factor in section~\ref{sec:hs_bias}. 
Note that in the limit of $N_{\rm draws}$ going to infinity, this equation becomes an exact representation of the integral in eq.~\ref{eq:likelihood}.
We have checked that further increasing the number of draws does not affect our result.
Also note that this expression corresponds to the first term in eq.~\ref{eq:joint_like}. We incorporate the second term (boost factor likelihood) as described in the following section.

\subsection{Sampling}

For each cluster bin in $z$ and $\tilde{y}_0$ we have six halo/scaling relation parameters ($A_{\rm mass}$, $\eta$, $\zeta$, $\alpha$, $\beta$ and $a_{\rm 2h}$), two boost factor parameters ($R_s$, $B_0$) and two miscentering parameters ($f_{\rm mis}$, $\sigma_{\rm mis}$). For the bias in the source redshift distribution and the multiplicative shear bias, we randomly select values from the Gaussian distributions described in section~\ref{sec:photoz_bias} and section~\ref{sec:shear_bias} per model calculation. The details of the free parameters are summarized in table~\ref{tab:parameters}.

\begin{table}
        \centering
	\begin{tabular}{ccc}
    \hline
            Parameter & Prior & Description \\ \hline
            $A_{\rm mass}$ & $\mathcal{U}$(0.1, 10.0) & Amplitude of \\ 
                           &                          & the hydrostatic mass bias \\ [2pt]
            $\eta$ & $\mathcal{U}$(-10, 10) &  Mass power-law of \\
                   &                       & the hydrostatic mass bias \\ [2pt]
            $\zeta$ & $\mathcal{U}$(-10, 10) & Redshift power-law of \\
                    &                       & the hydrostatic mass bias \\ [2pt]
            $\alpha$ & $\mathcal{U}$(0, 2) & Transition power-law of \\
                     &                     & the baryon profile \\ [2pt]
            $\beta$ & $\mathcal{U}$(2, 5) & Large-radius power-law of \\
                    & $\mathcal{U}$(2, 8) [wide] & the baryon profile \\ [2pt]
            $a_{\rm 2h}$ & $\mathcal{U}$ (0.01, 2.00) & Amplitude factor \\ 
                         &                            & for the 2-halo term \\ [2pt]
            $f_{\rm mis}$ & $\mathcal{N}$(0.18, $0.08^2$) & Fraction of \\
                         &                               & miscentered clusters \\ [2pt]
            $\ln \sigma_{\rm mis}$ [Mpc] & $\mathcal{N}$(-0.82, $0.22^2$) & Miscentering amplitude \\ [2pt]
            $R_s$ [Mpc] & $\mathcal{U}$(0.1, 7.0) & Scale radius of boost factor \\ [2pt]
            $B_0$ & $\mathcal{U}$(0.1, 7.0) & Amplitude of boost factor \\
    \hline
	\end{tabular}
    \caption{The description of the free parameters used in our analysis and the corresponding prior ranges. $\mathcal{U}(a,b)$ denotes a uniform prior with the lower (upper) bound of a (b), while $\mathcal{N}(\mu,\sigma^2)$ represents a Gaussian distribution with the mean $\mu$ and the standard deviation $\sigma$.}
\label{tab:parameters}
\end{table}

We use the Markov Chain Monte Carlo (MCMC) method to sample the posterior distribution of our model parameters, as implemented in the publicly available package, \textsc{emcee} \cite{emcee}. Since our cluster sample is similar to that of Shin et al. \cite{Shin2021} (the only difference is the SNR cut for the Compton-$y$ signal), we use their posterior for the miscentering parameters as our prior. 
Note that Kelly et al. \cite{Kelly2023}, using the DES-Y3 clusters constructed from the same dataset from which most of our BCGs are confirmed \cite{Hilton2021}, did not find any significant trend of miscentering parameters in richness (therefore in mass and SNR in Compton-$y$ signal; see their table 4). After converting the unit to what is used in this study, it is, $f_{\rm mis}=0.18\pm0.08$ and $\ln \sigma_{\rm mis} [{\rm Mpc}]=-0.82\pm0.22$. We have checked that changing this prior within a reasonable range does not meaningfully alter our main result.

To reduce the running time for MCMC chains to converge and to constrain boost factor parameters only from the boost factor measurements, we first run chains only for the boost factor parameters (the second term in eq.~\ref{eq:joint_like}). Then, when performing MCMC analyses for our main measurements, $\boldsymbol{g}_t$, we randomly sample a set of boost factor parameters from these chains to apply to the model. Note that this sampling method is equivalent to the importance sampling method. Mathematically, while our joint likelihood (eq.~\ref{eq:joint_like}) does not change its form under this sampling method, it does modify the evidence to
\begin{equation}
    \int d\theta_{\gamma} [\mathcal{P}(\hat{\boldsymbol{g}}_t|\theta_{\gamma},\theta_{\rm boost}, \{Y_{\rm obs}\}) \mathcal{P}(\theta_{\gamma})] \int d\theta_{\rm boost} [\mathcal{P} (\hat{\mathcal{B}}|\theta_{\rm boost}) \mathcal{P}(\theta_{\rm boost})].
\end{equation}
Note that the evidence for the fully joint analysis is,
\begin{equation}
    \iint d\theta_{\gamma} d\theta_{\rm boost} [\mathcal{P}(\hat{\boldsymbol{g}}_t|\theta_{\gamma},\theta_{\rm boost}, \{Y_{\rm obs}\}) \mathcal{P} (\hat{\mathcal{B}}|\theta_{\rm boost}) \mathcal{P}(\theta_{\gamma}) \mathcal{P}(\theta_{\rm boost})].
\end{equation}
That is, in our case, we split the posterior for $(\theta_{\gamma}, \theta_{\rm boost})$ into two separate posteriors:
\begin{equation}
    \mathcal{P} (\theta_{\gamma}, \theta_{\rm boost}|\hat{\boldsymbol{g}}_t, \hat{\mathcal{B}}) = \mathcal{P} (\theta_{\gamma}|\hat{\boldsymbol{g}}_t, \theta_{\rm boost}) \mathcal{P} (\theta_{\rm boost}|\hat{\mathcal{B}}),
\end{equation}
where the second term on the right-hand side is realized in a Monte Carlo integration manner by random sampling when marginalizing over the boost factor parameters. Given that the fitted boost factor models unbiasedly describe the original boost factor measurement, this method is equivalent to correcting the data first with the boost factor and then running shear likelihood chains.

To verify the convergence of our chains given the stochasticity in our likelihood, we use the rank-normalized split $\hat{R}$ diagnostic (the revised Gelman-Rubin $\hat{R}$ statistic in \cite{Vehtari2021}). In all chains, we find that all parameters satisfy $\hat{R}<1.05$ ($<1.03$ for the main parameters, $A_{\rm mass}$, $\eta$, $\zeta$), indicating stable sampling of the posterior.

\section{Results}
\label{sec:result}

\subsection{Hydrostatic mass bias without mass or redshift evolution}
\label{sec:result_no_evolution}

\begin{figure*}
    \centering
    \includegraphics[width=0.99\columnwidth]{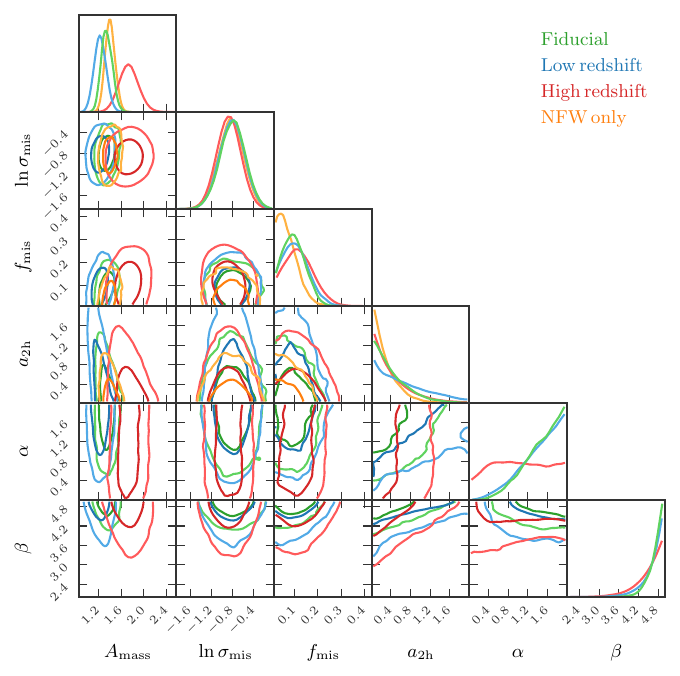}
    \caption{The posterior contours for the model parameters without mass and redshift evolution ($\eta=0$, $\zeta=0$). The green contours correspond to all cluster bins combined ($A_{\rm mass}=1/(1-b)=1.35^{+0.10}_{-0.10}$, $1-b = 0.74^{+0.06}_{-0.05}$), the blue contours the low redshift clusters ($z=[0.15,0.43]$, $A_{\rm mass}=1.23^{+0.11}_{-0.10}$, $1-b = 0.81^{+0.08}_{-0.06}$), and the red contours the high redshift clusters ($z=[0.43,0.7]$, $A_{\rm mass}=1.73^{+0.17}_{-0.16}$, $1-b = 0.58^{+0.06}_{-0.05}$). The orange contours are the results when an NFW-only halo model is adopted (no baryon correction, eq.~\ref{eq:baryon}; $A_{\rm mass}=1/(1-b)=1.41^{+0.08}_{-0.09}$, $1-b = 0.71^{+0.05}_{-0.04}$). See table~\ref{tab:parameters} for the description of the parameters.
    }
    \label{fig:posterior_zsplit}
\end{figure*}

In this section, we describe our result on constraining hydrostatic mass bias without mass or redshift evolution ($\eta=0$ and $\zeta=0$, eq.~\ref{Eq:Mtrue}). In addition, we report the results when each cluster redshift bin is separately used (the low-z sample, $z=[0.15,0.43]$ and the high-z sample, $z=[0.43, 0.70]$, respectively). 

In figure~\ref{fig:posterior_zsplit}, we show the posterior contours of our free parameters without mass or redshift evolution in green. The constrained 1-$\sigma$ range for $A_{\rm mass}$ is $1.35^{+0.10}_{-0.10}$, which translates to $1-b = 0.74^{+0.06}_{-0.05}$. Note that our main parameter $A_{\rm mass}$ does not show any significant degeneracy with any other parameters. We test the robustness of our result against possible systematic effects in section~\ref{sec:result_systematics}.

The posterior for $\ln \sigma_{\rm mis}$ ($-0.79^{+0.23}_{-0.24}$) is dominated by the prior ($-0.82 \pm 0.22$, table~\ref{tab:parameters}) and does not show any significant degeneracy with $A_{\rm mass}$. However, the posterior for $f_{\rm mis}$ prefers smaller values ($0.10^{+0.05}_{-0.05}$) than the prior ($0.18\pm0.08$) and show a weak degeneracy with $A_{\rm mass}$ (a larger miscentering fraction corresponding to a higher suppression of the $\boldsymbol{g}_t$ model at small scales, therefore a higher constrained mass). 
As for $a_{\rm 2h}$, since we have already chosen the radial ranges for the data vectors such that the 2-halo contribution is marginal (section~\ref{sec:measurement}), we prefer not to interpret the posterior that peaks at the lower bound of the prior. 
However, note that this behavior has been already shown in Cromer et al. \cite{Cromer2022} in which the constrained mass is less than 1\% biased, over the fitting range of 1-12 Mpc. Also, they found a consistent result on the mass constraint when excluding the 2-halo term from the model, fitting over the radii of 0.1 - 5.0 Mpc ($0.7^{+3.6}_{-4.5}\%$ mass bias with the 2-halo term, and $2.3^{+3.5}_{-4.9}\%$ mass bias without the 2-halo term), and that the 2-halo contribution is minimal (see their table 1 and figure 12). This result demonstrates the robustness of the mass constraints against inclusion of the two-halo term using our adopted model. However, given that the degeneracy of $a_{\rm 2h}$ with $A_{\rm mass}$ is non-negligible (a mild anti-correlation), we defer further tests of $a_{\rm 2h}$ to follow-up cosmological studies. Lastly, both $\alpha$ and $\beta$ peak at the upper bounds. Since $\alpha$ is the transition power law of the gNFW profile at $0.5 r_{\rm s}$ (section~\ref{sec:halo_model}) which is typically $\sim$0.1-0.15 $h^{-1}$Mpc, our measurement is not sensitive to it and we do not interpret the result. For $\beta$, a higher value corresponds to a sharper cutoff of the baryonic profile. We test in section~\ref{sec:result_systematics} with a wider prior for $\beta$, and confirm that it has a negligible effect on our mass constraint (table~\ref{tab:1_minus_b}) without any apparent degeneracy with $A_{\rm mass}$ (see figure~\ref{fig:posterior_mor}). Note that due to the weak degeneracies between these poorly constrained parameters and $A_{\rm mass}$, the posterior for $A_{\rm mass}$ is not significantly affected.

The $p$-value for the best-fit is $p=0.023$ ($\chi^2/{\rm dof} = 62.1/42$)\footnote{Here, we calculate the effective $\chi^2$ as $\chi_{\rm eff}^2=-2\times\ln[{\rm max(Likelihood)}$] from the chain. The degree of freedom (dof) is the number of data points (48 for the full analysis, 24 for the redshift-split analysis) minus the number of free parameters.}, hinting at possible discordance between the data and the model. Galaxy clusters at lower redshift are expected to be more relaxed (or thermalized) on average than those at higher redshift, due to a longer time since formation and/or mergers. Therefore, one would expect the hydrostatic mass bias to be smaller for the lower redshift clusters. Therefore, we perform a similar analysis (without mass and redshift evolution) on the low-z sample and the high-z sample separately. 

\begin{table}
        \centering
	\begin{tabular}{cc}
    \hline
        description & $1-b$ \\ \hline
        fiducial &  $0.74^{+0.06}_{-0.05}$ \\ [2pt]
        $z\in[0.15, 0.43]$ & $0.81^{+0.08}_{-0.06}$ \\ [2pt]
        $z\in[0.43, 0.70]$ & $0.58^{+0.06}_{-0.05}$ \\ [2pt] \hline
        without baryon modeling & $0.71^{+0.05}_{-0.04}$ \\[2pt]
        free $\sigma_{\ln M}$ (scatter in mass) & $0.76^{+0.06}_{-0.06}$ \\[2pt]
        wider $\beta$ prior (see table~\ref{tab:parameters}) & $0.76^{+0.07}_{-0.06}$ \\[2pt]
        $\times$5 wider $f_{\rm mis}$ and $\ln_{\rm mis}$ priors (miscentering) & $0.78^{+0.05}_{-0.07}$ \\[2pt]
        exclude first radial bin (scale cut) & $0.73^{+0.06}_{-0.05}$ \\
    \hline
	\end{tabular}
    \caption{The constrained values for $1-b$ without the mass and redshift evolution, including redshift split and systematics tests (section~\ref{sec:result_systematics}).
    }
\label{tab:1_minus_b}
\end{table}

The posterior contours for the analysis with the low-z (high-z) sample are shown in blue (red) in figure~\ref{fig:posterior_zsplit}. The constrained hydrostatic mass bias for the low-z sample is $A_{\rm mass} = 1.23^{+0.11}_{-0.10}$ or $1-b=0.81^{+0.08}_{-0.06}$, and that for the high-z sample is $A_{\rm mass} = 1.73^{+0.17}_{-0.16}$ or $1-b=0.58^{+0.06}_{-0.05}$. As expected, the low-z sample exhibits a smaller degree of hydrostatic mass bias ($\sim$18\% underestimation of mass) than the high-z sample ($\sim$42\% underestimation of mass). The difference between the high-z and low-z bias is $\sim$2.6$\sigma$. In table~\ref{tab:1_minus_b}, we summarize the results in this section (the first block of the table).

The $p$-value for the best fit is $p=0.014$ ($p=0.528$) for the low-z (high-z) sample with the $\chi^2/{\rm dof}$ of 33.5/18 (16.9/18). While the high-z sample exhibits a high $p$-value demonstrating that the model fits the data well, that of the low-z sample still shows a tension between the best-fit model and the data. 

These results motivate us to further investigate the redshift (and mass) trend of the hydrostatic mass bias. In the next section, we describe the result when the mass and the redshift evolution parameters ($\eta$, $\zeta$) are set free. Furthermore, in section~\ref{sec:result_systematics}, we show that including possible systematic effects (which could affect the $p$-value significantly) does not change our constraint on the hydrostatic mass bias significantly, demonstrating the robustness of our result. 

\subsection{Constraints on the redshift and mass evolution of hydrostatic mass bias}
\label{sec:result_evolution}

\begin{figure}
    \centering
    \includegraphics[width=0.99\columnwidth]{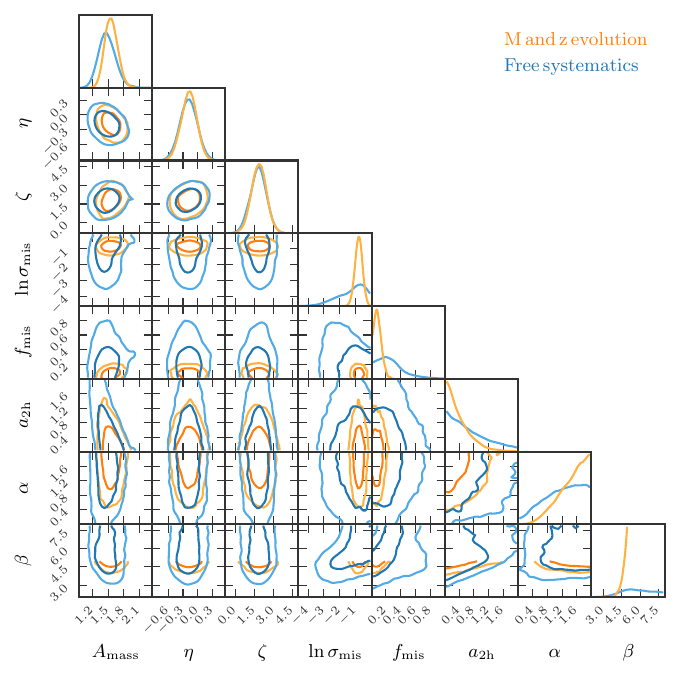}
    \caption{
    The posterior contours for the model parameters with mass and redshift evolution (free $\eta$ and $\zeta$). The orange contours correspond to our fiducial analysis (see section~\ref{sec:model} and \ref{sec:likelihood}), The blue contours when we restrict $A_{\rm mass}>1$, and the pink contours when including possible systematics: $5\times$ wider miscentering priors, a wider prior for $\beta$ (a baryonic parameter), exclusion of the first angular bin of each data vector (baryonic uncertainty), and setting $\sigma_{\ln M}$ free (instead of fixing at 0.2, section~\ref{sec:hs_bias}). Our fiducial analysis detects the evolution of hydrostatic mass bias in redshift ($\zeta>0$, decreasing hydrostatic mass bias with decreasing redshift) with 99.95\% confidence (3.3 $\sigma$), while the mass evolution parameter $\eta$ is only 1.2$\sigma$ away from being null detection.
    }
    \label{fig:posterior_mor}
\end{figure}

In figure~\ref{fig:posterior_mor}, we show the posterior contours of the parameters with the mass and redshift evolution ($\eta$, $\zeta$) in orange contours. First, we report that the hydrostatic mass bias at the pivot point ($z=0.45$, $M_{\rm SZ, 500c}=3 \times 10^{14} M_{\odot}$) is $1-b=0.64^{+0.06}_{-0.04}$, or $A_{\rm mass}=1.56^{+0.12}_{-0.12}$. The $p$-value for model fitting is $p=0.07$ ($\chi^2/{\rm dof} = 53.5/40)$. The 1$\sigma$ best-fit ranges for tangential shear measurements are shown as green shades in figure~\ref{fig:gammat_fit} and those for boost factor measurements in figure~\ref{fig:boost_fit} similarly. In red shades in figure~\ref{fig:gammat_fit}, we plot the 2-halo contribution from our model fitting, which confirms the minimal 2-halo contribution to the total model (see section~\ref{sec:result_no_evolution}; also see section~\ref{sec:result_systematics} for systematic tests regarding the 2-halo term). Also, we show the fractional residual of the tangential shear measurement with respect to the 68\% confidence range on the model fit in the bottom panel of figure~\ref{fig:gammat_fit}.

In particular, we find evidence for a strong dependence of the hydrostatic mass bias on redshift, with the redshift-dependence power-law of $\zeta=1.8^{+0.6}_{-0.5}$. Note that the corresponding power law for $1-b$ is $-\zeta$ (section~\ref{sec:hs_bias}). The corresponding significance of $\zeta>0$ is 99.95\% ($\sim$3.3$\sigma$). This increasing hydrostatic mass bias as a function of redshift is consistent with the current understanding of the dynamical evolution of halos: older halos (given the same current mass) are dynamically more relaxed, having had more time to thermalize, and therefore the measured hydrostatic mass is expected to be less biased than the younger component at an earlier epoch (higher $z$) \cite[e.g.,][]{Nelson2014, Shi2016}. 

On the other hand, our result does not show a clear trend of hydrostatic mass bias in the halo mass, with the probability of $\eta<0$ being only 88.11\% ($\sim$1.2$\sigma$), or $\eta=-0.18^{+0.15}_{-0.15}$. Therefore, we do not try to interpret this result and defer further analyses to follow-up studies.

\begin{figure*}
    \centering
    \includegraphics[width=0.99\columnwidth]{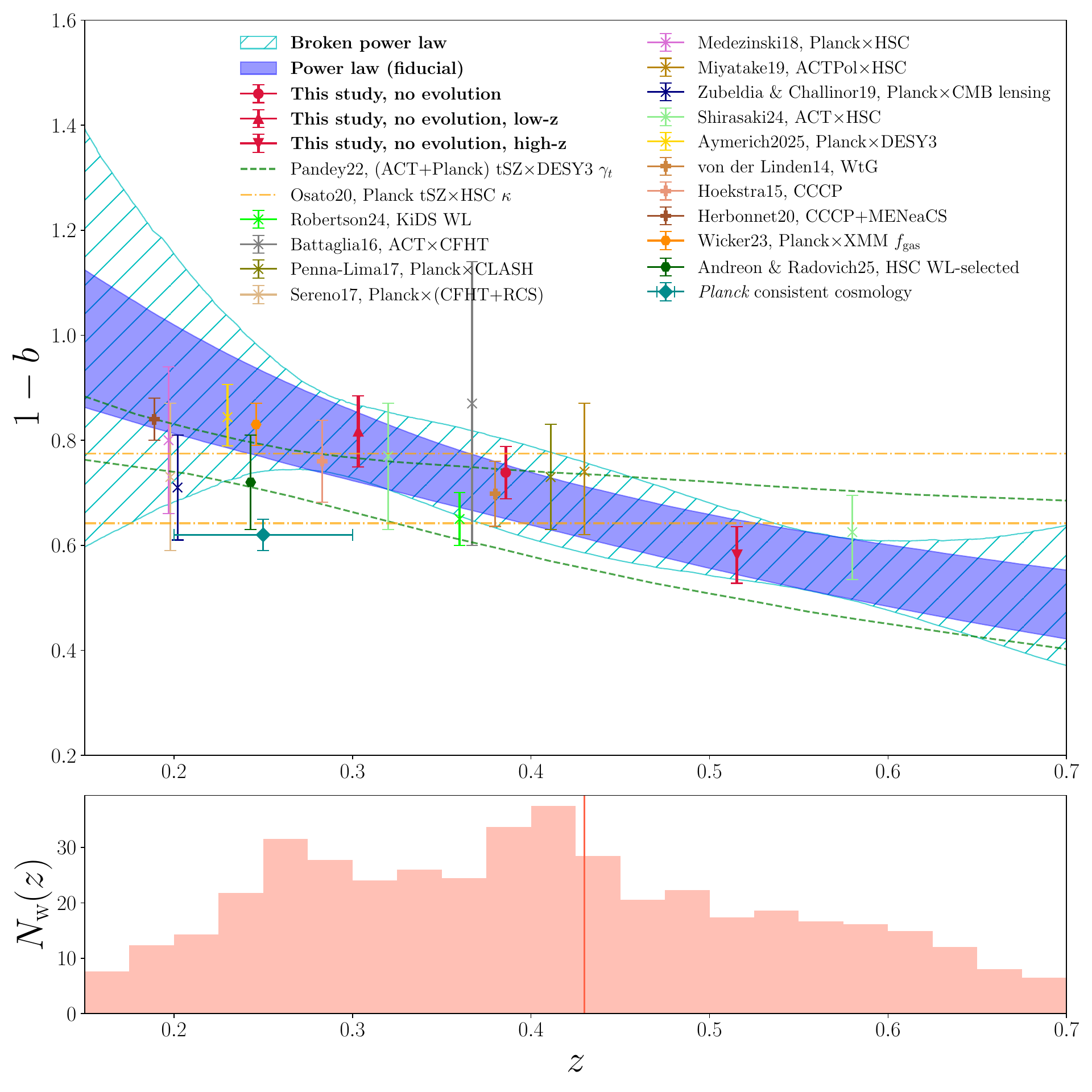}
    \caption{\textit{Top}: The redshift evolution of the hydrostatic mass bias ($1-b = 1/A_{\rm mass}$), where the SZ mass is fixed at $M_{\rm 500c}=3\times10^{14} M_{\odot}$. The blue band corresponds to the single power law model, and the cyan band corresponds to the broken power law model. 
    The red data points represent the results in this study with no mass or redshift evolution, for which the effective mean redshift values are calculated with the lensing weight applied. 
    Shown in cross ($\times$) markers are constrains from previous studies using tSZ clusters, and the plus markers correspond to those with X-ray clusters. 
    Also, the 1-$\sigma$ constraint from Pandey et al. \cite{Pandey2022} which used the cross-correlation between ACT+\textit{Planck} tSZ map and DES-Y3 shear is shown in dashed dark green lines. 
    Similarly, the result from Osato et al. \cite{Osato2020} using \textit{Planck} tSZ map cross-correlated with the HSC convergence field is shown in dot-dashed dark yellow lines. 
    The ``\textit{Planck} consistent cosmology'' corresponds to the hydrostatic mass bias required to exactly match their main CMB anisotropy cosmology ($\sigma_8$) with their cluster count cosmology result (note that, had we demanded the agreement to be within 2-$\sigma$, the required value for $1-b$ would have been larger, possibly agreeing with our fiducial result). 
    We refer the reader to section~\ref{sec:comparison} for further details of other data points here.\\
    \textit{Bottom}: The light red histogram shows the lensing-power-weighted distribution of the cluster redshift, normalized to be summed to the total number of clusters (443), where the vertical red line corresponds to the redshift split (section~\ref{sec:data_act}).
    }
    \label{fig:Amass_M_z}
\end{figure*}

In figure~\ref{fig:Amass_M_z}, we visualize the hydrostatic mass bias as a function of the redshift, at the pivot mass of $M_{\rm 500c, SZ} = 3 \times 10^{14} M_{\odot}$ (blue shade). Also plotted are the constraints on the hydrostatic mass bias from previous studies (cross markers correspond to SZ clusters, and plus markers to X-ray clusters). ``\textit{Planck} consistent cosmology'' corresponds to the hydrostatic mass bias needed to reconcile the fiducial \textit{Planck} 2018 cosmology \cite{Planck2018} with the cosmological constraints from the \textit{Planck} cluster count \cite{PlanckCluster2016}. We compare our result with other literature in more detail in section~\ref{sec:comparison}.

While our result is generally consistent with the earlier findings, the previous low redshift constraints ($z \lesssim 0.3$) hint at a possible flattening in the redshift trend. Therefore, we try a smoothly broken power law model for the redshift evolution: instead of $(\frac{1+z}{1+0.45})^\zeta$, using
\begin{eqnarray}
    \Big(\frac{1 + z}{1 + z_{\rm p}}\Big)^{\zeta_{\rm low}} \Bigl\{ \frac{1}{2} \Big[ 1 + \Big( \frac{1 + z}{1 + z_{\rm p}} \Big)^{1/\Delta_z} \Big] \Bigl\}^{(\zeta_{\rm high} - \zeta_{\rm low}) \Delta_z}.
\end{eqnarray}
Note that in this case, the pivot redshift, $z_{\rm p}$, is set free to capture the flattening redshift, the $\Delta_z$ is fixed at 0.05 and we have checked that this choice does not significantly affect our result as long as the transition is kept reasonably sharp. This profile follows the power law of $\zeta_{\rm low}$ at $z \lesssim z_{\rm p}$, and $\zeta_{\rm high}$ at $z \gtrsim z_{\rm p}$. For the newly added parameters, $\zeta_{\rm low}$, $\zeta_{\rm high}$ and $z_{\rm p}$, we use uniform priors of $\mathcal{U}(-10,10)$, $\mathcal{U}(-10,10)$ and $\mathcal{U}(0.15,0.7)$, respectively.

The result (cyan) is consistent with that from a power law (blue). However, we notice that when the broken power law is used, the value of the hydrostatic mass bias below $z\sim0.25$ becomes uncertain, while being marginally more consistent with the previous results than the power law case. Therefore, we conclude that while we detect the evolving hydrostatic mass bias as a function of redshift, the trend below $z\sim0.25$ remains uncertain. Note that the completeness of the ACT clusters sharply decreases below $z\sim0.2$ \cite[][figure 7]{Hilton2021}, which may partially explain this behavior.

To evaluate whether the additional complexity of the broken power law model could be justified by the data, we calculate the Akaike information criterion (AIC) and Bayesian information criterion (BIC) for the pure power law model and the broken power law model. The AIC and the BIC are derived as:
\begin{eqnarray}
    {\rm AIC} &=& 2k - 2\ln (\hat L) \\
    {\rm BIC} &=& k \ln (n) - 2\ln (\hat L),
\end{eqnarray}
where $k$ is the number of estimated parameters in the model, $\hat L$ is the maximum likelihood for the model, and $n$ is the number of data points. The smaller the value is, the more probable the model is to be an optimal model. Note that they penalize adding new parameters in the model. If $\Delta$ is the difference in the AIC or BIC between two models, the model with the smaller value is $\exp(|\Delta|/2)$ times more likely to be an optimal model.

For both AIC and BIC, our data prefers the pure power law model ($\Delta = 4.64$ and $8.38$, respectively, corresponding to 10 and 66 times larger probability). Thus we find that the additional complexity of the broken power law is not statistically warranted, and we retain the single power law as our fiducial model.

\subsection{Systematics tests}
\label{sec:result_systematics}

We first test the effect of modeling baryons in our halo model (eq.~\ref{eq:baryon}). We run another MCMC chain without modeling baryons (an NFW-only model). The result is shown in the orange contours in figure~\ref{fig:posterior_zsplit}. As a result, we obtain $A_{\rm mass} = 1.41^{+0.08}_{-0.09}$, or $1-b = 0.71^{+0.05}_{-0.04}$. Compared with our fiducial result with baryon modeling, $A_{\rm mass} = 1.35^{+0.10}_{-0.10}$, not modeling baryon results in a $\sim$3.2-5.3\% higher mass estimate (assuming a perfect correlation between the results), although being well within 1-$\sigma$. This result is consistent within $\sim$1$\sigma$ with the bias reported in Cromer et al. \cite{Cromer2022} (a $+7.5 \pm 1.9$\% bias). Note that the robustness of our halo model against possible mass bias due to halo model mismatch relies on the analysis of Cromer et al. \cite{Cromer2022}, therefore, the shown consistency between ours and Cromer et al. \cite{Cromer2022} result is reassuring.

Our analysis choice (section~\ref{sec:model} and \ref{sec:likelihood}) involves several components that are potentially susceptible to systematic effects. In this section, we test whether or not our modeling choice would be robust against these possible systematics. We first test four effects as follows. 
\begin{itemize}
    \item Set $\sigma_{\ln M}$ (log-normal scatter in the mass) free between 0.05 and 0.5 instead of fixing it at 0.2 (no baryon case).
    \item Relax the flat prior range for $\beta$ (the long-distance power-law of the baryon profile) from $\mathcal{U}(2,5)$ to $\mathcal{U}(2,8)$ (the ``wide'' prior in table~\ref{tab:parameters}). Note that since $\alpha$ is the transition power-law of baryons which applies around $0.5 r_{\rm s}$ ($\sim$0.15$h^{-1}{\rm Mpc}$ for our cluster sample) that is comparable to or smaller than our minimum angular scale cut, we do not include this parameter here, since it would not change our model for tangential shear significantly according to eq.~\ref{eq:dsigma} (wide baryon prior case).\footnote{Also note that according to figure~\ref{fig:posterior_zsplit}, $A_{\rm mass}$ and $\alpha$ do not show any meaningful correlation. In addition, the next systematic test below excluding the first radial bin, where our minimum scale cut becomes much larger than the scale where $\alpha$ matters, shows a shift in $A_{\rm mass}$ much smaller than the uncertainty.}
    \item Amplify the standard deviations of the priors for the miscentering parameters ($\ln \sigma_{\rm mis}$ and $f_{\rm mis}$) 5 times (significantly flatter priors; wide miscentering prior case).
    \item Exclude the first radial bins of the data vectors from the fitting (scale cut case).
\end{itemize}

We first apply these systematic changes one by one to our fiducial analysis with no mass and redshift evolution (section~\ref{sec:result_no_evolution}). The resultant constraints from each of these four cases are summarized in the second block of table~\ref{tab:1_minus_b}. The constrained $1-b$ values from the respective four cases are $0.76^{+0.06}_{-0.06}$, $0.76^{+0.07}_{-0.06}$, $0.78^{+0.05}_{-0.07}$ and $0.73^{+0.06}_{-0.05}$, all of which are almost the same as or well within 1-$\sigma$ from our fiducial result of $0.74^{+0.06}_{-0.05}$, demonstrating the robustness of our constraint on $1-b$ against these possible systematics.

We then perform an MCMC analysis again with these modifications all at once with the mass and redshift evolution. The result is shown in figure~\ref{fig:posterior_mor} with blue contours (``Free systematics''). The best-fit $p$-value for this analysis is 0.13 ($\chi^2/{\rm dof}=40.1/31$, since we have an additional parameter for the scatter in the mass, and excluded the first radial bins) which is much higher than 0.07 of the fiducial result. Despite this fact, we confirm that the constraints of the hydrostatic mass bias parameters ($1.48^{+0.16}_{-0.16}$, $-0.19^{+0.17}_{-0.17}$, and $1.8^{+0.6}_{-0.6}$, for $A_{\rm mass}$, $\eta$, and $\zeta$ respectively) from this analysis are well within 1-$\sigma$ (within 0.1$\sigma$ for $\eta$ and $\zeta$) of our fiducial result (orange, also see table~\ref{tab:Nconverge}) while the constraints of the other parameters have changed significantly, demonstrating the robustness of our result on the hydrostatic mass bias constraint against the possible systematics listed above.

\begin{table}
        \centering
	\begin{tabular}{cccc}
    \hline
        description & $A_{\rm mass}$ & $\eta$ & $\zeta$\\ \hline
        fiducial, no evolution &  $1.35^{+0.10}_{-0.10}$ & - & - \\ [2pt]
        $N_{\rm draws} = 13$, no evolution & $1.35^{+0.10}_{-0.09}$ & - & - \\ [2pt]
        $N_{\rm draws} = 20$, no evolution & $1.35^{+0.10}_{-0.09}$ & - & - \\ [2pt]
        different seed, no evolution & $1.35^{+0.09}_{-0.10}$ & - & - \\ [2pt]
        \hline
        fiducial, with evolution & $1.56^{+0.12}_{-0.12}$ & $-0.18^{+0.15}_{-0.15}$ & $1.8^{+0.6}_{-0.5}$\\ [2pt]
        $N_{\rm draws} = 13$, with evolution & $1.56^{+0.12}_{-0.12}$ & $-0.18^{+0.16}_{-0.16}$ & $1.8^{+0.6}_{-0.5}$ \\ [2pt]
        $N_{\rm draws} = 20$, with evolution & $1.56^{+0.12}_{-0.12}$ & $-0.18^{+0.15}_{-0.16}$ & $1.8^{+0.6}_{-0.5}$ \\ [2pt]
        different seed, with evolution & $1.56^{+0.12}_{-0.13}$ & $-0.18^{+0.16}_{-0.15}$ & $1.9^{+0.5}_{-0.6}$ \\ [2pt]
        \hline
        excluding the last radial bin & $1.63^{+0.15}_{-0.14}$ & $-0.28^{+0.17}_{-0.16}$ & $1.9^{+0.6}_{-0.6}$ \\[2pt]
        excluding $\tilde{y}_0 \in$ [0.00,1.05] $\times$ $z \in$ [0.43,0.70]* & $1.63^{+0.17}_{-0.18}$ & $-0.24^{+0.18}_{-0.18}$ & $2.1^{+0.7}_{-0.7}$ \\[2pt]
        redshift-dependent miscentering & $1.58^{+0.12}_{-0.12}$ & $-0.18^{+0.15}_{-0.16}$ & $1.8^{+0.6}_{-0.6}$ \\[2pt]
        $A_{\rm mass} > 1$ & $1.50^{+0.16}_{-0.16}$ & $-0.23^{+0.25}_{-0.26}$ & $2.3^{+0.9}_{-0.9}$ \\ [2pt]
        \hline
	\end{tabular}
    \caption{The constrained values for $A_{\rm mass}$, $\eta$ and $\zeta$, when varying $N_{\rm draws}$ from 7 to 13 and 20 and when using a different random seed for the stochasticity, as well as the constraints from additional systematic tests to the mass/redshift evolution (see section~\ref{sec:result_systematics}). First block: with no mass/redshift evolution, second block: with mass/redshift evolution. Third block: additional systematics test. \\ *: the cluster bin with the marginal null PTE (see section~\ref{sec:measurement_null}).
    }
\label{tab:Nconverge}
\end{table}

In addition, for the Monte Carlo (MC) integral of the likelihood (eq.~\ref{eq:likelihood} to \ref{eq:likelihood_approx}), we used $N_{\rm draws}=7$. To test the convergence of our stochastic chain with $N_{\rm draws}=7$, we run another MCMC chain for the model without mass and redshift evolution (section~\ref{sec:result_no_evolution}) with $N_{\rm draws}=13$ and 20, keeping everything else identical. For our key parameter $A_{\rm mass}$, we obtain $1.35^{+0.10}_{-0.09}$ ($1.35^{+0.10}_{-0.09}$) for $N_{\rm draws}=13$ (20), compared with $1.35^{+0.10}_{-0.10}$ for the $N_{\rm draws}=7$ case (well within 0.1$\sigma$ shift in the central value and in the upper and lower bound). Also, to test the stability of our result against the adopted random seed for the stochasticity, we run a $N_{\rm draws}=7$ chain with a different random seed, obtaining $A_{\rm mass} = 1.35^{+0.09}_{-0.10}$, well within 0.05$\sigma$ shift in the central value and in the upper and lower bound. The results are summarized in the first block of table~\ref{tab:Nconverge}. Finally, we repeat these tests ($N_{\rm draws}=13$, $20$ and using a different random seed) for the case with the mass and redshift evolution. For $N_{\rm draws}=13$ (20), we find $A_{\rm mass}=1.56^{+0.12}_{-0.12}$ ($1.56^{+0.12}_{-0.12}$), $\eta=-0.18^{+0.16}_{-0.16}$ ($-0.18^{+0.15}_{-0.16}$), and $\zeta=1.8^{+0.6}_{-0.5}$ ($1.8^{+0.6}_{-0.5}$). For the run with a different seed, we find $A_{\rm mass}=1.56^{+0.12}_{-0.13}$, $\eta=-0.18^{+0.16}_{-0.15}$, and $\zeta=1.9^{+0.5}_{-0.6}$. We confirm that all of these results are within 0.05$\sigma$ from our fiducial result and we summarize them in the second block of table~\ref{tab:Nconverge}. Overall, these results demonstrate that the numerical noise from our MC likelihood integral is subdominant to the statistical uncertainty of the data itself. Consequently, we conclude that the choice of $N_{\rm draws}=7$ provides a sufficient level of precision for robust parameter inference without introducing any significant numerical bias.

Furthermore, we perform four additional systematics tests to ensure the robustness of the constraints of $A_{\rm mass}$, $\eta$, and $\zeta$ as follows:
\begin{itemize}
    \item Excluding the last radial bins to test the sensitivity to the 2-halo contribution.
    \item Excluding the cluster bin of $\tilde{y}_0 \in$ [0.00,1.05] and $z \in$ [0.43,0.70] to test the robustness of the parameter constraints against the marginal PTE for the null test (table~\ref{tab:null_chisq}).
    \item Assuming different sets of miscentering parameters ($\ln \sigma_{\rm mis}$, $f_{\rm mis}$) for different redshift bins to test the sensitivity to the evolving miscentering parameters.
    \item Limiting the minimum value of $1/(1-b)$ at 1 (a physical prior).
\end{itemize}

First, when excluding the last radial bin from the fitting, we obtain $A_{\rm mass}=1.63^{+0.15}_{-0.14}$, $\eta=-0.28^{+0.17}_{-0.16}$, and $\zeta=1.9^{+0.6}_{-0.6}$. Compared with the fiducial value of $A_{\rm mass}=1.56^{+0.12}_{-0.12}$, $\eta=-0.18^{+0.15}_{-0.15}$, and $\zeta=1.8^{+0.6}_{-0.5}$, we confirm the shifts well within 1$\sigma$. The non-zero detection significance for $\zeta$ remains strong, at 99.90\% confidence level (3.1$\sigma$), while that of $\eta$ remains at a similar level ($95.51\%$, 1.7$\sigma$). This result demonstrates the robustness of our key parameter constraints against the 2-halo contribution.

Excluding the cluster bin with the marginal null PTE ($\tilde{y}_0 \in$ [0.00,1.05], $z \in$ [0.43,0.70]), the constraints for $A_{\rm mass}$, $\eta$ and $\zeta$ are $1.63^{+0.17}_{-0.18}$, $-0.24^{+0.18}_{-0.18}$, and $2.1^{+0.7}_{-0.7}$, respectively. These values are well within 1$\sigma$ of those from the fiducial constraints and confirm that this cluster bin does not meaningfully bias our conclusion.

Next, assuming different sets of miscentering parameters for the low-z and high-z samples results in $A_{\rm mass}=1.58^{+0.12}_{-0.12}$, $\eta=-0.18^{+0.15}_{-0.16}$, and $\zeta=1.8^{+0.6}_{-0.6}$, almost the same as the fiducial result. This result along with the previous test with a wider miscentering prior tells us that miscentering systematics are not likely to significantly affect our fiducial results.

Finally, when forcing the value of $1/(1-b)$ to be larger or equal than 1, we obtain $A_{\rm mass} = 1.50^{+0.16}_{-0.16}$, $\eta = -0.23^{+0.25}_{-0.26}$, and $\zeta = 2.3^{+0.9}_{-0.9}$. Again, these constraints are well within 1$\sigma$ ranges of our fiducial results. The results for these additional systematics tests are summarized in the third block of table~\ref{tab:Nconverge}.

Lastly, to test the robustness of our redshift evolution result against the selection of prior, we applied the importance reweighting technique to our MCMC posterior chain with the mass and redshift evolution (section~\ref{sec:result_evolution}). The importance reweighting is a technique to account for the change from the original prior to a different prior that one wants to test, by applying the weight, $p_{\rm new} (\theta) / p_{\rm old} (\theta)$, to each sample, where $\theta$ is the parameter in the sample, $p_{\rm old}$ the original prior and $p_{\rm new}$ the prior in test. In this case, we would like to test if our redshift evolution parameter constraint ($\zeta = 1.8^{+0.6}_{-0.5}$, $\sim$3.3$\sigma$) is robust against a prior that prefers no evolution, therefore, we use $p_{\rm new} = \mathcal{N} (0, 1^2)$. When applying this new prior with the importance reweighting to our chain, we still obtain $\zeta=1.4^{+0.5}_{-0.5}$ ($p(\zeta>0) = 0.998$, $\sim$2.9$\sigma$ significance). The effective sample size ratio (ESS/N = 0.59) confirms that the reweighting is statistically reliable. Thus, we conclude that our detection of the redshift evolution in the hydrostatic mass bias is data-driven and robust against prior choices. Note that even when we use $p_{\rm new} = \mathcal{N} (0, 0.7^2)$, $\zeta=1.1^{+0.5}_{-0.4}$ ($p(\zeta>0) = 0.995$, $\sim$2.6$\sigma$ significance, ESS/N = 0.28).

\subsection{Comparison to other literature}
\label{sec:comparison}

In this section, we compare our constraint on the hydrostatic mass bias to the results from other literature with various probes. 

First, we compare our result to the previous literature with SZ-selected clusters. 
In Robertson et al. \cite{Robertson2024}, they used the ACT DR5 clusters (the same as ours) that overlap with the Kilo Degree Survey (KiDS, 157 clusters), obtaining the WL constraint for $1-b$ of $0.65\pm0.05$ at the effective redshift of 0.36. This result is about 1$\sigma$ lower than our result at the corresponding redshift (see figure~\ref{fig:Amass_M_z}). 
In Battaglia et al. \cite{Battaglia2016}, using nine clusters from early ACT data and WL from the Canada-France-Hawaii telescope (CFHT) survey, they derived $1-b=0.87\pm0.27$ at $z\sim0.37$. 
Also, in Penna-Lima et al. \cite{PennaLima2017}, using 21 clusters in common between the \textit{Planck} and the Cluster Lensing And Supernova survey with Hubble (CLASH), they obtained $1-b=0.73\pm0.10$ at $z\sim0.4$. 
Sereno et al. \cite{Sereno2017}, using 35 \textit{Planck} clusters covered by CFHT and the Red-sequence Cluster Survey (RCS2), reported $1-b = 0.73 \pm 0.11 \, {\rm (stat)} \pm 0.08 \, {\rm (sys)}$ at $z\sim0.2$. 
Meanwhile, in Medezinski et al. \cite{Medezinski2018}, using five \textit{Planck} clusters that overlap with the Subaru Hyper Suprime-Cam (HSC) survey, they obtained $1-b = 0.80 \pm 0.14$ at $z\sim0.2$.
More recently, Miyatake et al. \cite{Miyatake2019} reported $1-b = 0.74^{+0.13}_{-0.12}$ using eight ACT clusters that overlap with the HSC first-year WL data, at the effective redshift of 0.43.
In addition, calibrating the cluster masses through the CMB lensing using \textit{Planck} data, Zubeldia \& Challinor \cite{Zubeldia2019} obtained $1-b = 0.71 \pm 0.10$ at $z\sim0.2$. 
Shirasaki et al. \cite{Shirasaki2024} used 96 ACT DR5 clusters overlapping with the HSC Year 3 footprint, divided into three redshift bins of $z=[0.092,0.445]$, $[0.445, 0.695]$ and $[0.695,1.180]$, finding the WL constraints of $1/(1-b) = 1.3\pm0.2$, $1.6\pm0.2$ and $1.6\pm0.3$, respectively. Note that this result of Shirasaki et al. \cite{Shirasaki2024} hints at a possible redshift evolution of hydrostatic mass bias as found in this study.
Recently, Aymerich et al. \cite{Aymerich2025} used 93 \textit{Planck} tSZ clusters that overlap with the DESY3 footprint, and the WL shape catalog from the DESY3 (the same as ours), found that $1-b = 0.844^{+0.055}_{-0.062}$.
These constraints on hydrostatic mass bias are generally in good agreement with our result. We overplotted these results on top of ours in figure~\ref{fig:Amass_M_z}, with the ``cross ($\times$)'' markers.

In von der Linden et al. \cite{vonderLinden2014}, using 38 X-ray clusters from the Weighing the Giants survey which overlap with the \textit{Planck} clusters, they derived the WL constraint of $1-b = 0.698 \pm 0.062$ at $z=0.38$. 
Also, Hoekstra et al. \cite{Hoekstra2015}, as part of the Canadian Cluster Comparison Project (CCCP), reported $1-b = 0.76 \pm 0.05 \, {\rm (stat)} \pm 0.06 \, {\rm (sys)}$, for 50 X-ray selected clusters through WL analysis with various sources at $z\sim0.28$.
In addition, including $\sim$50 additional clusters from the Multi Epoch Nearby Cluster Survey (MENeaCS), Herbonnet et al. \cite{Herbonnet2020} obtained WL constraints of $1-b = 0.84 \pm 0.04 \, {\rm (stat)} \pm 0.05 \, {\rm (sys)}$ at $z\sim0.19$. 
These results related to X-ray clusters are plotted with the ``plus $(+)$'' markers in figure~\ref{fig:Amass_M_z}, which are in good agreement with our result as well. 

On the other hand, cross-correlation measurements between tSZ and WL shear (or convergence) provide an independent method to constrain hydrostatic mass bias, through its leverage on both gas properties and total matter profile. 
Cross-correlating the WL shear from the DES Y3 (the same WL data as this study) with the tSZ measurement from ACT+\textit{Planck}, Pandey et al. \cite{Pandey2022} reported the hydrostatic mass bias as a function of redshift, shown in green dashed lines (1$\sigma$ range), by which we confirm the agreement between their result and ours.
Moreover, in Osato et al. \cite{Osato2020}, they cross-correlated the WL convergence map ($\kappa$) from the HSC first-year data and the tSZ map from \textit{Planck}2015 \cite{Planck2016tSZ}, reporting $1-b = 0.731^{+0.044}_{-0.089}$, which also agrees well with our result. 
The agreement between our result from the cluster WL method and theirs from tSZ$\times$WL cross-correlations implicitly demonstrates the robustness of our result. 

Meanwhile, in Wicker et al. \cite{Wicker2023}, deriving the hydrostatic mass bias anchored on the gas mass fraction from the follow-up X-ray observations from XMM within 120 massive galaxy clusters from \textit{Planck}, which is expected to follow the global baryon fraction in the Universe $\Omega_{\rm b}/\Omega_{\rm m}$, they have found $1-b = 0.83 \pm 0.04$ at the mean redshift of 0.246. Notably, they have found a significant detection of the redshift evolution of the mass bias, $\zeta=0.64 \pm 0.18$, which agrees with our result in sign but is somewhat discrepant quantitatively ($\sim$2$\sigma$). Also note that Andreon \& Radovich \cite{Andreon2025}, using 13 gravity-selected clusters from the HSC data and the Compton-$y$ values from ACT, derived $1-b = 0.72 \pm 0.09$.

\textit{Planck}2018 \cite{Planck2018} reports the hydrostatic mass bias required to reconcile their main cosmological constraints from the CMB anisotropies with that from the cluster number count \cite{PlanckCluster2016,PlanckClusters}, $1-b = 0.62 \pm 0.03$ (also see ref. \cite{Salvati2019}). Given that the effective average redshift of the \textit{Planck} clusters is 0.2-0.3, it is more than 2$\sigma$ discrepancy from our result (see figure~\ref{fig:Amass_M_z}, the diamond marker). Note, however, that the recent study of Lee et al. \cite{Lee2024} hints at a discrepancy in nature between the hydrostatic mass bias from \textit{Planck} and that from ACT, that $1-b$ from the ACT data is larger than that from the Planck data in their joint analysis, attributing it to possible redshift evolution or the expansion history of the Universe.

\section{Discussions and conclusion}
\label{sec:discussion}

In this study, we have presented a galaxy weak-lensing mass calibration analysis of the SZ-selected clusters from the ACT DR5 data using the galaxy weak-lensing shear catalog constructed from the first three years of the Dark Energy Survey data. 

Our data vectors (figure~\ref{fig:gammat_fit} and \ref{fig:boost_fit}) comprise the tangential shear and the boost factor as a function of radial angular distance, both of which are calculated within two cluster redshift bins and two cluster mass (Compton-$y$) bins. The choice of tangential shear as a function of angular distance (instead of, for example, excess surface density $\Delta\Sigma$ as a function of physical distance) is to avoid any cosmological dependence of the data vectors themselves.
The boost factor (contamination from cluster member galaxies to the shear catalog) is calculated using the $P(z)$ decomposition method \cite{Gruen2014,Varga2019}, with a slightly improved fitting treatment (incorporating the bootstrapped uncertainty of $P(z)$ instead of a simple least-squares algorithm).
The covariance matrices of the data vectors are calculated using the bootstrap method with $N=100,000$.

We then fully forward-model the (reduced) tangential shear around our cluster sample, starting from the three-component halo model: an NFW profile (1-halo dark matter distribution), a generalized NFW profile (1-halo baryon distribution) and a 2-halo profile. As shown in Cromer et al. \cite{Cromer2022}, this three-component halo model accurately constrains the weak-lensing halo mass within 1\% bias, so that we avoid calibrating the mass bias induced by a weak-lensing analysis using hydrodynamical simulations. When converting this 3-dimensional halo density profile into the observed (reduced) tangential shear, we take into account various systematics: uncertainty on the distribution of the source redshift, uncertainty on the multiplicative shear bias, miscentering of clusters, and the boost factor. Note that we do not correct our data vector with the measured boost factor, but instead correct our tangential shear model with the parameterized representation of the boost factor within the model to fully forward-model our data vector. 

We assume the hydrostatic mass bias has power law dependencies on both redshift and mass. We also test the case where the hydrostatic mass bias is constant, as well as one where the bias has a broken power law dependence on redshift.

Also, instead of using a halo mass function or assuming a single halo profile to fit the stacked weak-lensing profile, we first model individual clusters with the corresponding masses drawn from the underlying mass-observable relation \cite{Hilton2021} with a scatter in the mass-observable relation along with the hydrostatic mass bias applied. Then we stack the individual cluster models to generate the prediction for the final stacked tangential shear profiles.

We then perform MCMC analyses to constrain the hydrostatic mass bias parameters. To reduce computation time, we first constrain the boost factor parameters ($2\times$[2 source bins]$\times$[4 cluster bins]$=$16 parameters) and then randomly sample from this boost factor chain when running our main chain for the tangential shear fitting. 

Since we separately model each individual cluster and then calculate the stacked tangential shear model thereof, our likelihood is intrinsically an N-dimensional integral (N being the number of clusters). To avoid a considerable amount of computation time, we approximate this N-dimensional integral in cluster masses with the average likelihood of stochastic draws of cluster masses from an underlying distribution, with $N_{\rm draws}=7$. Note that while our approximation is adequate to asymptote to the true solution in an MCMC chain, as $N_{\rm draws} \to \infty$, it becomes analytically accurate.

When the hydrostatic mass bias is fixed in mass and redshift (no evolution), we find $1-b=0.74^{+0.06}_{-0.05}$. This value is generally in agreement with previous results ($\sim$0.65-0.85). When the low-redshift and high-redshift cluster bins are fitted separately, we report the hydrostatic mass bias values of $0.81^{+0.08}_{-0.06}$ and $0.58^{+0.06}_{-0.05}$, respectively, showing a clear redshift trend. 

We then free the mass and redshift evolution parameters. We find an increasing trend of the hydrostatic mass bias in redshift, with 99.95\% significance (power law of $1.8^{+0.6}_{-0.5}$), while we do not detect any significant trend in cluster mass. Although our result as a function of redshift is largely consistent with previous studies, there is a hint of flattening of the trend at low redshift below $\sim$0.3. To capture this possible flattening trend and also to give the model more freedom, we have applied a broken power law model to the redshift evolution of hydrostatic mass bias. The resultant trend agrees with the single power law case very well above $z\sim0.25$, although that below $z\sim0.25$ becomes highly uncertain.

In particular, cross-correlation analyses between tSZ and optical weak-lensing provide an independent probe to study the redshift trend of hydrostatic mass bias \cite{Osato2020,Pandey2022}. Our result is in excellent agreement with these results. 

The increasing trend of the hydrostatic mass bias with increasing redshift has also been shown in previous simulation-based studies as increasing non-thermal pressure support with redshift \cite[e.g.,][]{Nelson2014, Shi2016}, which is largely modulated by the mass accretion rate into clusters. While our result is qualitatively consistent with these studies, whether this trend would continue at higher redshift ($>0.7$) remains unclear, depending on thermal and mass accretion history of galaxy clusters. We expect that a combination of next-generation cosmological surveys such as the WL analysis with the Legacy Survey of Space and Time (LSST) of Rubin Observatory using the tSZ cluster sample from Simons Observatory (SO) would enable further detailed studies of mass and redshift trends of hydrostatic mass bias with much improved precision and extended mass/redshift ranges.

There are a few remaining caveats to address in future studies. First of all, we used the miscentering posterior from Shin et al. \cite{Shin2021} as the prior in our analysis. However, while our redshift selection is consistent with theirs, we apply a higher cut on the SNR of Compton-$y$ parameter (5.5 instead of 4). In principle, this can affect the degree of miscentering of the selected clusters. While we have tested this effect by applying five times wider miscentering prior, we defer further evaluation of the miscentering distribution of our cluster sample to future studies (e.g., using eROSITA X-ray clusters; ref. \cite{Bulbul2024}). Also note that Currie et al. \cite{Currie2025} recently demonstrated that the miscentering parameters of galaxy clusters could be more precisely constrained using the Gaussian mixture model approach, where the final likelihood comprises the multiplication of likelihoods from individual clusters with a mixture of Gaussian probabilities of well-centered and miscentered contribution. 

While the priors or set values for the parameters in our gNFW baryon density model (eq.~\ref{eq:baryon}) are conditioned on the simulation that Cromer et al. \cite{Cromer2022} used, the baryon profiles from a simulation could differ from what we actually observe. Although this model has been widely used in previous studies, especially those studying baryonic feedback \cite{Schneider2019,Giri2021,Grandis2024,Bigwood2024,Dalal2025}, a general consensus on the values of these parameters in massive clusters has not yet been reached. However, note that we have demonstrated that increasing the prior range for the baryonic parameter $\beta$ (the large-scale power law; our data vector is not sensitive to $\alpha$ and $\gamma$ due to the much larger scale used in the measurements than where $\alpha$ and $\gamma$ matter) does not significantly alter our conclusion. 

Also, in the future, we expect to be able to perform a joint study of WL mass calibration and baryonic feedback by including X-ray and tSZ observations (gas profiles) to simultaneously constrain the baryonic parameters and the mass of clusters (see, e.g., references above and refs. \cite{Gatti2022,Pandey2022} for techniques to constrain baryonic feedback within clusters), or obtain reasonable prior ranges from various hydrodynamical simulations.

Shin et al. \cite{Shin2021} showed that the clusters used in this study demonstrate excellent agreement in shape between the total matter profile determined by WL and the galaxy surface density profile from the cluster-galaxy correlation, given the cut in galaxy magnitude ($M_i>-19.87$). In follow-up analyses, we would explore the possibility of combining the WL data with the galaxy density profile to have better constraints on the shape of the total matter field, which in turn could enhance the constraint on the cluster mass. 

In this analysis, we have used data-driven covariance matrices. However, a data-driven covariance matrix is inherently a noisy and sometimes biased realization of the true underlying covariance, which affects the parameter constraints in the corresponding MCMC analysis. In principle, one could overcome much of this difficulty by using an analytically calculated covariance matrix \cite[e.g.,][]{Wu2019}. However, deriving an accurate analytical covariance matrix requires careful consideration of all non-linear effects that could contribute to the cluster profiles. This is beyond the scope of this paper, and we defer further analyses on this regard to future studies.   

Lastly, while we adopt a hierarchical model in that we model individual clusters and then stack them, our Bayesian formalism is not fully hierarchical (as it does not use individual WL profiles of clusters, but instead a stacked data vector). Although it is possible to use a fully hierarchical Bayesian framework \cite[e.g.,][]{Bocquet2019,Bocquet2023,Bocquet2024,Grandis2024lensing} to improve the precision of the mass constraint, modeling individual clusters involves much more complicated systematics than a stacked weak-lensing analysis, such as triaxiality, halo orientation bias as well as halo-to-halo variations on, e.g., large-scale projection, miscentering and concentration \cite[e.g.,][]{Mandelbaum2008,Becker2011,Herbonnet2019,Herbonnet2022}. In the future, we plan to combine our analysis with the cluster number count cosmology pipeline \cite[e.g.,][]{Lee2024}, to simultaneously obtain constraints on cosmological parameters such as $\Omega_{\rm m}$ and $\sigma_8$, with an optimal methodology that we will explore.

\acknowledgments

Funding for the DES Projects has been provided by the U.S. Department of Energy, the U.S. National Science Foundation, the Ministry of Science and Education of Spain, 
the Science and Technology Facilities Council of the United Kingdom, the Higher Education Funding Council for England, the National Center for Supercomputing 
Applications at the University of Illinois at Urbana-Champaign, the Kavli Institute of Cosmological Physics at the University of Chicago, 
the Center for Cosmology and Astro-Particle Physics at the Ohio State University,
the Mitchell Institute for Fundamental Physics and Astronomy at Texas A\&M University, Financiadora de Estudos e Projetos, 
Funda{\c c}{\~a}o Carlos Chagas Filho de Amparo {\`a} Pesquisa do Estado do Rio de Janeiro, Conselho Nacional de Desenvolvimento Cient{\'i}fico e Tecnol{\'o}gico and 
the Minist{\'e}rio da Ci{\^e}ncia, Tecnologia e Inova{\c c}{\~a}o, the Deutsche Forschungsgemeinschaft and the Collaborating Institutions in the Dark Energy Survey. 

The Collaborating Institutions are Argonne National Laboratory, the University of California at Santa Cruz, the University of Cambridge, Centro de Investigaciones Energ{\'e}ticas, 
Medioambientales y Tecnol{\'o}gicas-Madrid, the University of Chicago, University College London, the DES-Brazil Consortium, the University of Edinburgh, 
the Eidgen{\"o}ssische Technische Hochschule (ETH) Z{\"u}rich, 
Fermi National Accelerator Laboratory, the University of Illinois at Urbana-Champaign, the Institut de Ci{\`e}ncies de l'Espai (IEEC/CSIC), 
the Institut de F{\'i}sica d'Altes Energies, Lawrence Berkeley National Laboratory, the Ludwig-Maximilians Universit{\"a}t M{\"u}nchen and the associated Excellence Cluster Universe, 
the University of /, NSF NOIRLab, the University of Nottingham, The Ohio State University, the University of Pennsylvania, the University of Portsmouth, 
SLAC National Accelerator Laboratory, Stanford University, the University of Sussex, Texas A\&M University, and the OzDES Membership Consortium.

Based in part on observations at NSF Cerro Tololo Inter-American Observatory at NSF NOIRLab (NOIRLab Prop. ID 2012B-0001; PI: J. Frieman), which is managed by the Association of Universities for Research in Astronomy (AURA) under a cooperative agreement with the National Science Foundation.

The DES data management system is supported by the National Science Foundation under Grant Numbers AST-1138766 and AST-1536171.
The DES participants from Spanish institutions are partially supported by MICINN under grants PID2021-123012, PID2021-128989 PID2022-141079, SEV-2016-0588, CEX2020-001058-M and CEX2020-001007-S, some of which include ERDF funds from the European Union. IFAE is partially funded by the CERCA program of the Generalitat de Catalunya.

We acknowledge support from the Brazilian Instituto Nacional de Ci\^encia 
e Tecnologia (INCT) do e-Universo (CNPq grant 465376/2014-2).

This document was prepared by the DES Collaboration using the resources of the Fermi National Accelerator Laboratory (Fermilab), a U.S. Department of Energy, Office of Science, Office of High Energy Physics HEP User Facility. Fermilab is managed by Fermi Forward Discovery Group, LLC, acting under Contract No. 89243024CSC000002.

Support for ACT was through the U.S. National Science Foundation through awards AST-0408698, AST-0965625, and AST-1440226 for the ACT project, as well as awards PHY-0355328, PHY-0855887 and PHY-1214379. Funding was also provided by Princeton University, the University of Pennsylvania, and a Canada Foundation for Innovation (CFI) award to UBC. ACT operated in the Parque Astron\'omico Atacama in northern Chile under the auspices of the Agencia Nacional de Investigaci\'on y Desarrollo (ANID). The development of multichroic detectors and lenses was supported by NASA grants NNX13AE56G and NNX14AB58G. Detector research at NIST was supported by the NIST Innovations in Measurement Science program. Computing for ACT was performed using the Princeton Research Computing resources at Princeton University, the National Energy Research Scientific Computing Center (NERSC), and the Niagara supercomputer at the SciNet HPC Consortium. SciNet is funded by the CFI under the auspices of Compute Canada, the Government of Ontario, the Ontario Research Fund–Research Excellence, and the University of Toronto. We thank the Republic of Chile for hosting ACT in the northern Atacama, and the local indigenous Licanantay communities whom we follow in observing and learning from the night sky.

TS was supported by a grant from the Simons Foundation (Simons Investigator in Astrophysics, Award ID 620789). AvdL and TS acknowledge support by grants from the US Department of Energy (SC0018053 and SC0023387). NB acknowledges support for this work from NASA grant 80NSSC22K0410 and additional support from NASA grant 80NSSC22K0721. KM acknowledges support from the National Research Foundation of South Africa. MH acknowledges support from the National Research Foundation of South Africa (grant nos. 97792, 137975, CPRR240513218388). TM acknowledges support from the Agencia Estatal de Investigaci\'on (AEI) and the Ministerio de Ciencia, Innovaci\'on y Universidades (MICIU) Grant ATRAE2024-154740 funded by MICIU/AEI//10.13039/501100011033. AN acknowledges support from the European Research Council (ERC) under the European Union’s Horizon 2020 research and innovation program with Grant agreement No. 101163128. JRB acknowledges support from NSERC. JD acknowledges NSF grant AST-2108126. CS acknowledges support from the Agencia Nacional de Investigaci\'on y Desarrollo (ANID) through Basal project FB210003.

\bibliographystyle{JHEP}
\bibliography{biblio}

\end{document}